\documentclass[10pt,letterpaper,english,aps,prl,reprint,twocolumn]{revtex4}
\usepackage[latin9]{inputenc}
\usepackage{babel}
\usepackage{verbatim}
\usepackage{amsbsy}
\usepackage{amstext}
\usepackage{amssymb}
\usepackage{graphicx}
\usepackage{esint}
\usepackage[unicode=true,
 bookmarks=true,bookmarksnumbered=false,bookmarksopen=false,
 breaklinks=false,pdfborder={0 0 1},backref=false,colorlinks=false]
 {hyperref}
\hypersetup{pdftitle={Non-linear optical response of non-interacting systems with spectral methods},
 pdfauthor={S. M. João, J. M. V. P. Lopes},
 pdfsubject={physics, article, article_personal},
 pdfkeywords={Nonlinear Optics, Keldysh, KPM, Kernel Polynomial Method, Condensed Matter Physics}}

\makeatletter

\pdfpageheight\paperheight
\pdfpagewidth\paperwidth

\@ifundefined{textcolor}{}
{%
 \definecolor{BLACK}{gray}{0}
 \definecolor{WHITE}{gray}{1}
 \definecolor{RED}{rgb}{1,0,0}
 \definecolor{GREEN}{rgb}{0,1,0}
 \definecolor{BLUE}{rgb}{0,0,1}
 \definecolor{CYAN}{cmyk}{1,0,0,0}
 \definecolor{MAGENTA}{cmyk}{0,1,0,0}
 \definecolor{YELLOW}{cmyk}{0,0,1,0}
}

\makeatother

\begin{document}
\global\long\def\ket#1{\left|#1\right\rangle }%

\global\long\def\bra#1{\left\langle #1\right|}%

\global\long\def\braket#1#2{\left.\left\langle #1\right|#2\right\rangle }%

\title{Basis-Independent Spectral Methods for Non-linear Optical Response
in Arbitrary Tight-binding Models}
\author{S. M. João}
\affiliation{Centro de Física das Universidades do Minho e Porto }
\affiliation{Departamento de Física e Astronomia, Faculdade de Ciências, Universidade
do Porto, 4169-007 Porto, Portugal}
\author{J. M. V. P. Lopes}
\affiliation{Centro de Física das Universidades do Minho e Porto }
\affiliation{Departamento de Engenharia Física, Faculdade de Engenharia}
\affiliation{Departamento de Física e Astronomia, Faculdade de Ciências, Universidade
do Porto, 4169-007 Porto, Portugal}
\begin{abstract}
In this paper, we developed a basis-independent perturbative method
for calculating the non-linear optical response of arbitrary non-interacting
tight-binding models. Our method is based on the non-equilibrium Keldysh
formalism and allows an efficient numerical implementation within
the framework of the Kernel Polynomial Method for systems which are
not required to be translation-invariant. Some proof-of-concept results
of the second-order optical conductivity are presented for the special
case of gapped graphene with vacancies and an on-site Anderson disordered
potential.
\end{abstract}
\maketitle

\section{Introduction}

Since the advent of the laser in 1960, the field of non-linear optics
has received considerable interest. The following year marked the
beginning of a systematic study of this field, as in 1961 P. Franken
was able to demonstrate second harmonic generation (SHG) \citep{PhysRevLett.7.118}
experimentally. This opened the gateway to a whole new plethora of
phenomena, such as optical rectification (1962) \citep{bass1962optical}
and higher harmonic generation (1967) \citep{new1967optical}. The
laser was able to provide the powerful electric fields needed to see
these nonlinear phenomena. The optical properties of crystals have
been studied extensively throughout the last 30 years \citep{PhysRevB.48.11705}
and have recently received considerable interest due to the strong
non-linear properties of layered materials such as graphene \citep{0953-8984-20-38-384204,GLAZOV2014101,bonaccorsoGraphenePhotonicsOptoelectronics2010}.

Several approaches have been developed to obtain the non-linear response
up to arbitrary order of a crystalline system subject to an external
field. Among those, we may find generalizations of Kubo's formula
for higher orders and perturbation expansions for the density matrix
in both the length gauge and the velocity gauge \citep{PhysRevB.96.035431,PhysRevB.48.11705}.
Due to the complexity of the problem, the bulk of that work was done
using translation-invariant systems. Despite covering a broad range
of systems, this approach is limited in its scope, as it cannot deal
with impurities, border effects, or even magnetic fields within the
usual Peierls framework \citep{peierls1933theorie}. In order to simulate
more realistic systems, this restriction has to be lifted. The goal
is then to re-express the same formulas for the linear and nonlinear
optical conductivities in a framework that allows real-space calculations.
In this work, we extend the work of Passos et al \citep{passosNonlinearOpticalResponses2018}
implementing the velocity gauge methodology within the Keldysh formalism
\citep{keldysh1964diagram}. This way, we develop a general perturbation
procedure to deal with non-interacting fermion systems at finite temperature
coupled to a time-dependent external field. In \citep{weisse2004chebyshev},
an advanced Chebyshev expansion method is proposed to compute linear
response functions. This paper comes as the next logical step, by
providing the generalization to all orders in perturbation theory.

Through careful categorization of all these contributions, we provide
a systematic procedure to find the objects needed to calculate the
conductivity at any order. These objects are expressed with no reference
to a specific basis. The critical point here is that the mathematical
objects provided by our perturbation expansion are precisely the ones
required by the numerical spectral methods we use. This fact, combined
with our diagrammatic approach, provides a straightforward way to
implement the numerical calculation of the nonlinear optical conductivity
for a wide range of materials. This methodology is implemented in
KITE, an open-source software developed by ourselves \citep{quantumkite}
to calculate the nonlinear optical properties of a very broad range
of tight-binding models.

This paper is structured as follows. In \hyperref[Section II]{Section II},
we introduce the Keldysh formalism to show how the current will be
calculated and define the fundamental mathematical objects of our
calculations. \hyperref[Section III]{Section III} applies the perturbation
procedure to a tight-binding model. The electromagnetic field is added
through the Peierls Substitution. Then, a diagrammatic procedure is
developed in order to deal with the large number of non-trivial contributions
to the conductivity in each order. \hyperref[Section IV]{Section IV}
explains how to use a Chebyshev expansion of the operators in the
previous expression in order to be used with spectral methods. This
provides the full formula that may be directly implemented numerically.
Finally, in \hyperref[Section V]{Section V} we apply the formalism
developed in the previous sections to several different systems. We
start by comparing with known results obtained by $\boldsymbol{k}$-space
integration \citep{PhysRevB.96.035431,passosNonlinearOpticalResponses2018}
of the nonlinear optical conductivity of gapped graphene. We focus
solely on the intrinsic contribution of the crystal to the optical
conductivity, disregarding any excitonic effects. Then, we add some
disorder to the system and show that our method gives the expected
result. Finally, the numerical efficiency and convergence of our method
is assessed. 

\section{Keldysh Formalism\label{Section II}}

The Keldysh formalism \citep{keldysh1964diagram} is a general perturbation
scheme describing the quantum mechanical time evolution of non-equilibrium
interacting systems at finite temperature. It provides a concise diagrammatic
representation of the average values of quantum operators. This formalism
does not rely on any particular basis, which is a critical feature
for this paper. The mathematical objects obtained in our expansion
are in the form required by numerical spectral methods, providing
a straightforward formula for the numerical calculation of nonlinear
optical response functions. In this section we will introduce the
definitions of the objects used throughout the paper and show how
to expand the Green's functions for fermions \citep{jishi2013feynman}
with this formalism.

\subsection{Definitions}

\subsubsection{Green's functions}

To use the Keldysh formalism for fermions, we need the definitions
of the time-ordered, lesser, greater and anti-time-ordered Green's
functions. Respectively,

\begin{eqnarray}
iG_{ab}^{T}\left(t,t'\right) & = & \left\langle T\left[c_{a}\left(t\right)c_{b}^{\dagger}\left(t'\right)\right]\right\rangle \label{eq:Green's functions definition}\\
iG_{ab}^{<}\left(t,t'\right) & = & -\left\langle c_{b}^{\dagger}\left(t'\right)c_{a}\left(t\right)\right\rangle \\
iG_{ab}^{>}\left(t,t'\right) & = & \left\langle c_{a}\left(t\right)c_{b}^{\dagger}\left(t'\right)\right\rangle \\
iG_{ab}^{\tilde{T}}\left(t,t'\right) & = & \left\langle \widetilde{T}\left[c_{a}\left(t\right)c_{b}^{\dagger}\left(t'\right)\right]\right\rangle .
\end{eqnarray}

All the creation and annihilation operators are in the Heisenberg
picture and the labels $a$ and $b$ denote states belonging to a
complete single-particle basis. $T$ is the time-ordering operator
and $\tilde{T}$ the anti-time-ordering operator. The average $\left\langle \cdots\right\rangle $
stands for $\mbox{Tr}\left[\rho(t_{0})\cdots\right]/\mbox{Tr}\left[\rho(t_{0})\right]$
in the grand canonical ensemble, $\rho$ is the density matrix and
$t_{0}$ denotes the time at which the external perturbation has been
switched on. These are the building blocks of the Keldysh formalism.
The advanced and retarded Green's functions are a simple combination
of the previous objects:
\begin{eqnarray}
G^{R} & = & G^{T}-G^{<}\label{eq:Relation RT}\\
G^{A} & = & -G^{\tilde{T}}+G^{<}.\label{eq:relation AT}
\end{eqnarray}

The non-perturbed versions of these Green's functions are denoted
by a lowercase $g$.

\subsubsection{Expected value of an operator}

The expected value of the current $\boldsymbol{J}\left(t\right)$
(or any one-particle operator) may be evaluated with resort to these
Green's functions by tracing over its product with the perturbed lesser
Green's function:

\begin{equation}
\boldsymbol{J}\left(t\right)=\left\langle \hat{\boldsymbol{J}}\left(t\right)\right\rangle =-\text{Tr}\left[\hat{\boldsymbol{J}}\left(t\right)iG^{<}\left(t,t\right)\right].\label{current in terms of G<}
\end{equation}

\begin{figure}
\includegraphics{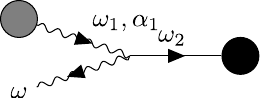}

\caption{Diagrammatic representation of the expected value of the current operator
in Fourier space. The horizontal straight line ending in a circle
is the lesser Green's function and the wavy line beginning in a circle
represents the current operator. }

\label{current operator}
\end{figure}

The Fourier transform \footnote{ Fourier convention: $f\left(t\right)=\left(2\pi\right)^{-1}\int d\omega e^{-i\omega t}\tilde{f}\left(\omega\right)$. }
of $\boldsymbol{J}\left(t\right)$ is shown diagrammatically in Figure
\ref{current operator}. The circles stand for the full, perturbed
operators in the presence of an external field.

\subsubsection{Conductivity}

We use the same definition for the nonlinear optical conductivity
as in \citep{PhysRevB.96.035431,passosNonlinearOpticalResponses2018}:

\begin{eqnarray}
 &  & J^{\alpha}\left(\omega\right)=\sigma^{\alpha\beta}\left(\omega\right)E^{\beta}\left(\omega\right)+\int\frac{\text{d}\omega_{1}}{2\pi}\int\frac{\text{d}\omega_{2}}{2\pi}\times\label{conductivity definition}\\
 &  & \sigma^{\alpha\beta\gamma}\left(\omega_{1},\omega_{2}\right)E^{\beta}\left(\omega_{1}\right)E^{\gamma}\left(\omega_{2}\right)2\pi\delta\left(\omega_{1}+\omega_{2}-\omega\right)+\cdots\nonumber 
\end{eqnarray}
where $E^{\alpha}$ is the component of the electric field along the
$\alpha$ direction and the repeated indices are assumed to be summed
over. The coefficients of this expansion are the conductivities at
each order in the expansion. The next section is devoted to finding
the perturbation expansion of $G^{<}$. In this paper we are dealing
with tight-binding models, in which case the current operator will
itself be a power series of the external field.

\subsection{Non-interacting electronic systems}

Our system is described by the many-particle time-dependent Hamiltonian

\begin{equation}
\mathcal{H}\left(t\right)=\mathcal{H}_{0}+\mathcal{H}_{\text{ext}}\left(t\right).
\end{equation}
where $\mathcal{H}_{0}$ is an Hamiltonian that we can solve exactly
and $\mathcal{H}_{\text{ext}}\left(t\right)$ is the time-dependent
external perturbation. Here we restrict ourselves to non-interacting
Hamiltonians since we're dealing with non-interacting electrons. These
operators are expressed in terms of their single-particle counterparts
as 
\begin{equation}
\mathcal{H}_{\text{ext}}\left(t\right)=\sum_{ab}\left[H_{\text{ext}}\left(t\right)\right]_{ab}c_{a}^{\dagger}\left(t\right)c_{b}\left(t\right)
\end{equation}

\begin{equation}
\mathcal{H}_{0}=\sum_{ab}\left[H_{0}\right]_{ab}c_{a}^{\dagger}\left(t\right)c_{b}\left(t\right).
\end{equation}

The expansion of the perturbed lesser Green's function $G^{<}$ will
be expressed in terms of the unperturbed Green's functions $g^{>}$,
$g^{<}$, $g^{R}$ and $g^{A}$ in Fourier space:

\begin{eqnarray}
i\tilde{g}^{<}\left(\omega\right) & = & -2\pi f\left(\hbar\omega\right)\delta\left(\omega-H_{0}/\hbar\right)\label{eq: lesser green's function}\\
i\tilde{g}^{>}\left(\omega\right) & = & 2\pi\left[1-f\left(\hbar\omega\right)\right]\delta\left(\omega-H_{0}/\hbar\right)\\
i\tilde{g}^{R}\left(\omega\right) & = & \frac{i}{\omega-H_{0}/\hbar+i0^{+}}\\
i\tilde{g}^{A}\left(\omega\right) & = & \frac{i}{\omega-H_{0}/\hbar-i0^{+}},
\end{eqnarray}
where $f\left(\epsilon\right)=\left(1+e^{\beta(\epsilon-\mu)}\right)^{-1}$
is the Fermi-Dirac distribution, $\beta$ is the inverse temperature
and $\mu$ is the chemical potential. The Keldysh formalism and Langreth's
rules provide the perturbation expansion of $G^{<}$ \citep{devreeseLinearNonlinearResponse1975}.
Defining $V\left(t\right)=\left(i\hbar\right)^{-1}H_{\text{ext}}\left(t\right)$,
the zeroth-order term in the expansion is

\begin{equation}
i\tilde{G}^{<\left(0\right)}(\omega)=\int\text{d}\omega_{1}i\tilde{g}^{<}(\omega_{1})\delta(\omega)\label{eq:green zeroth order frequencies}
\end{equation}
and the first-order one is

\begin{eqnarray}
 &  & i\tilde{G}^{<\left(1\right)}(\omega)=\int\frac{\text{d}^{3}\omega_{123}}{\left(2\pi\right)^{3}}\left(2\pi\right)^{2}\delta(\omega_{1}-\omega_{2}-\omega_{3})\label{eq:green first order frequencies}\\
 &  & \times\delta(\omega+\omega_{3}-\omega_{1})\left[i\tilde{g}^{R}(\omega_{1})\tilde{V}(\omega_{2})i\tilde{g}^{<}(\omega_{3})+\right.\nonumber \\
 &  & \left.+i\tilde{g}^{<}(\omega_{1})\tilde{V}(\omega_{2})i\tilde{g}^{A}(\omega_{3})\right].\nonumber 
\end{eqnarray}
$\int\text{d}^{n}\omega_{1\cdots n}$ is a shorthand for $\int\cdots\int\text{d}\omega_{1}\cdots\text{d}\omega_{n}$.
The second-order term is

\begin{eqnarray}
 &  & i\tilde{G}^{<(2)}(\omega)=\int\frac{\text{d}^{5}\omega_{1\cdots5}}{\left(2\pi\right)^{5}}\left(2\pi\right)^{3}\delta(\omega_{5}+\omega-\omega_{1})\times\nonumber \\
 &  & \delta(\omega_{1}-\omega_{2}-\omega_{3})\delta(\omega_{3}-\omega_{4}-\omega_{5})\times\nonumber \\
 &  & \left[i\tilde{g}^{R}(\omega_{1})\tilde{V}(\omega_{2})i\tilde{g}^{R}(\omega_{3})\tilde{V}(\omega_{4})i\tilde{g}^{<}(\omega_{5})\right.\nonumber \\
 &  & +i\tilde{g}^{R}(\omega_{1})\tilde{V}(\omega_{2})i\tilde{g}^{<}(\omega_{3})\tilde{V}(\omega_{4})i\tilde{g}^{A}(\omega_{5})\label{eq:second_order_green_frequencies-1}\\
 &  & \left.+i\tilde{g}^{<}(\omega_{1})\tilde{V}(\omega_{2})i\tilde{g}^{A}(\omega_{3})\tilde{V}(\omega_{4})i\tilde{g}^{A}(\omega_{5})\right].\nonumber 
\end{eqnarray}

\begin{figure}
\includegraphics[scale=0.5]{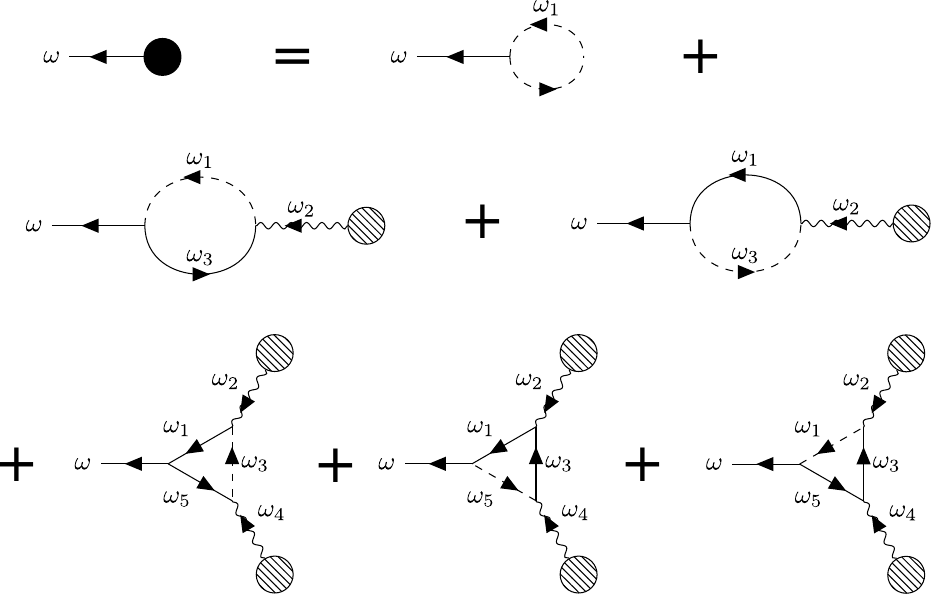}\caption{Diagrammatic representation of the lesser Green's function.}
\label{Diagrammatic representation of the lesser Green's function}

\end{figure}

Diagrammatically, the expansion of $iG^{<}\left(\omega\right)$ is
represented by Figure \ref{Diagrammatic representation of the lesser Green's function}.
Each wavy line ending in a circle represents an external perturbation
$\tilde{V}$. There are three different types of Green's functions
that may appear in these expansions, with a certain regularity: a
lesser Green's function $\tilde{g}^{<}$, which is always present,
retarded Green's functions $\tilde{g}^{R}$ and advanced Green's functions
$\tilde{g}^{A}$. Diagrammatically, $\tilde{g}^{<}$ is represented
by a dashed line while the solid lines represent retarded or advanced
Green's functions. To identify whether a line represents a retarded
or advanced Green's function, one needs to read the diagram and identify
the position of the lesser Green's function and the outgoing line.
Reading clockwise (anti-clockwise) until finding the outgoing line,
there can only be advanced (retarded) Green's functions. In each intersection,
the corresponding external perturbation $\tilde{V}$ is inserted.
An exception is made for the intersection with the line representing
$\omega$, as it still needs to be contracted.

If the external perturbation were a simple external field $\boldsymbol{E}\left(t\right)$,
then the coupling would be $H_{\text{ext}}\left(t\right)=e\boldsymbol{E}\left(t\right)\cdot\boldsymbol{r}$
and the previous expressions coupled with eq. (\ref{current in terms of G<})
would suffice. Now we will turn to tight-binding Hamiltonians, for
which the external coupling is actually an infinite series of operators
due to the way the electromagnetic field is introduced. This affects
not only the $V$ operators but also the expression for the current
operator.

\section{tight-binding Hamiltonian with external electric field\label{Section III}}

Tight-binding models provide a simple framework with which to calculate
transport quantities. This framework can be used to express structural
disorder in the system, whilst Peierls' substitution \citep{peierls1933theorie}
adds an electromagnetic field as an external perturbation. Despite
the simplicity of this procedure, the addition of an electromagnetic
field through a phase factor yields an infinite series of $H_{\text{ext}}$.
In this section, we obtain the expression for $H_{\text{ext}}$ and
show how the expansions of the previous sections may be used to obtain
the nonlinear optical conductivity. This is entirely analogous to
the way the external perturbation is introduced with the velocity
gauge in the work of Passos et al \citep{passosNonlinearOpticalResponses2018}.

\subsection{Series expansion}

Let's consider the following tight-binding Hamiltonian:

\begin{equation}
\mathcal{H}_{0}=\sum_{\boldsymbol{R}_{i},\boldsymbol{R}_{j}}\sum_{\sigma_{1},\sigma_{2}}t_{\sigma_{1}\sigma_{2}}\left(\boldsymbol{R}_{i},\boldsymbol{R}_{j}\right)c_{\sigma_{1}}^{\dagger}\left(\boldsymbol{R}_{i}\right)c_{\sigma_{2}}\left(\boldsymbol{R}_{j}\right).\label{eq:TB hamiltonian}
\end{equation}
The $\boldsymbol{R}_{i}$ represent the lattice sites and the $\sigma_{i}$
the other degrees of freedom unrelated to the position, such as the
orbitals and spin. The electromagnetic field is introduced through
Peierls' substitution:

\begin{equation}
t_{\sigma_{1}\sigma_{2}}\left(\boldsymbol{R}_{i},\boldsymbol{R}_{j}\right)\rightarrow e^{\frac{-ie}{\hbar}\int_{\boldsymbol{R}_{j}}^{\boldsymbol{R}_{i}}\boldsymbol{A}(\boldsymbol{r}',t)\cdot d\boldsymbol{r}'}t_{\sigma_{1}\sigma_{2}}\left(\boldsymbol{R}_{i},\boldsymbol{R}_{j}\right).\label{eq:peierls substitution}
\end{equation}

To introduce both a static magnetic field and a uniform electric field,
we use the following vector potential:

\begin{equation}
\boldsymbol{A}(\boldsymbol{r},t)=\boldsymbol{A}_{1}(\boldsymbol{r})+\boldsymbol{A}_{2}(t).
\end{equation}

The electric and magnetic fields are obtained from $\boldsymbol{E}(t)=-\partial_{t}\boldsymbol{A}_{2}(t)$
and $\boldsymbol{B}(\boldsymbol{r})=\boldsymbol{\nabla}\times\boldsymbol{A}_{1}(\boldsymbol{r})$.
The introduction of the magnetic field only changes the $t_{\sigma_{1}\sigma_{2}}\left(\boldsymbol{R}_{i},\boldsymbol{R}_{j}\right)$
without introducing a time dependency. Therefore, we may assume that
a magnetic field is always present without any loss of generality
for the following discussion while keeping in mind that its introduction
broke translation invariance. Since the magnetic field only affects
the hopping parameters, from now on, the term in the vector potential
that provides the electric field will be denoted by $\boldsymbol{A}\left(t\right)$.
The external perturbation is obtained by expanding the exponential
in eq. \ref{eq:peierls substitution} and identifying the original
Hamiltonian.

\subsubsection{Expansion of the external perturbation}

Expanding the exponential in eq. \ref{eq:peierls substitution} yields
an infinite series of operators for the full Hamiltonian:

\begin{equation}
H_{\boldsymbol{A}}\left(t\right)=H_{0}+H_{\text{ext}}\left(t\right)
\end{equation}
from which we identify, after a Fourier transform,
\begin{eqnarray}
 &  & \tilde{V}\left(\omega\right)=\frac{e}{i\hbar}h^{\alpha}\tilde{A}^{\alpha}\left(\omega\right)+\frac{e^{2}}{i\hbar}\frac{h^{\alpha\beta}}{2!}\int\frac{\text{d}\omega'}{2\pi}\int\frac{\text{d}\omega''}{2\pi}\times\quad\label{external perturbation}\\
 &  & \tilde{A}^{\alpha}\left(\omega'\right)\tilde{A}^{\beta}\left(\omega''\right)2\pi\delta\left(\omega'+\omega''-\omega\right)+\cdots.\nonumber 
\end{eqnarray}

Repeated spatial indices are understood to be summed over. We have
defined

\begin{equation}
\hat{h}^{\alpha_{1}\cdots\alpha_{n}}=\frac{1}{\left(i\hbar\right)^{n}}\left[\hat{r}^{\alpha_{1}},\left[\cdots\left[\hat{r}^{\alpha_{n}},H_{0}\right]\right]\right]\label{eq:generalized velocity}
\end{equation}
where $\hat{\boldsymbol{r}}$ is the position operator. In first order,
$\hat{h}^{\alpha}$ is just the single-particle velocity operator.
Under PBC, the position operator $\hat{\boldsymbol{r}}$ is ill-defined
but its commutator with the Hamiltonian is not. In real space, this
commutator is simply the Hamiltonian matrix element connecting the
two sites $i$ and $j$ multiplied by the distance vector $\boldsymbol{d}_{ij}$
between them. If we define this distance vector as the distance between
neighbors instead of the difference of the two positions, it will
be well defined in PBC. Using this strategy, all the $\hat{h}$ operators
may be evaluated in position space by assigning to each bond the Hamiltonian
matrix element multiplied by the required product of difference vectors
$h_{ij}^{\alpha_{1}\cdots\alpha_{n}}=\left(i\hbar\right)^{-n}H_{ij}d_{ij}^{\alpha_{1}}\cdots d_{ij}^{\alpha_{n}}$. 

\begin{figure}
\includegraphics[scale=0.6]{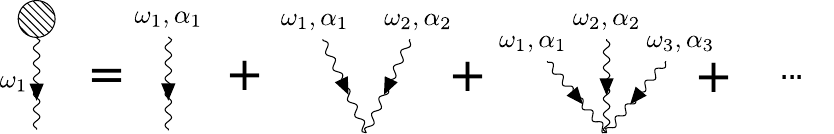}

\caption{Diagrammatic representation of the external perturbation.}
\label{Diagrammatic representation of the external perturbation}

\end{figure}

In Figure \ref{Diagrammatic representation of the external perturbation},
we see how the diagrammatic representation of the external perturbation
unfolds into an infinite series of external fields. The wavy line
represents $\tilde{\boldsymbol{A}}$ and the number of external fields
connected to the same point is the number of commutators in eq. \ref{eq:generalized velocity}.

\subsubsection{Expansion of the current}

The current operator is calculated directly from the Hamiltonian,
using $\hat{J}^{\alpha}=-\Omega^{-1}\partial H/\partial A^{\alpha}$
($\Omega$ is the volume of the sample), which also follows a series
expansion due to the presence of an infinite number of $\boldsymbol{A}\left(t\right)$
in $H_{\text{ext}}$:

\begin{eqnarray}
 &  & \hat{J}^{\alpha}\left(t\right)=-\frac{e}{\Omega}\left(\hat{h}^{\alpha}+e\hat{h}^{\alpha\beta}A^{\beta}\left(t\right)+\right.\nonumber \\
 &  & \left.\quad\quad+\frac{e^{2}}{2!}\hat{h}^{\alpha\beta\gamma}A^{\beta}\left(t\right)A^{\gamma}\left(t\right)+\cdots\right).\label{eq: current expansion}
\end{eqnarray}

\begin{figure}
\includegraphics[scale=0.7]{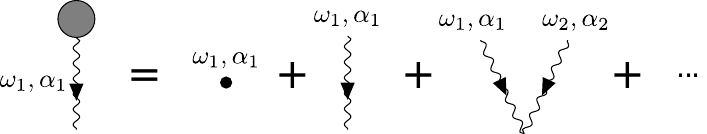}

\caption{Diagrammatic representation of the current operator. The single small
circle is to be understood as a Dirac delta.}
\label{Diagrammatic representation of the current operator.}
\end{figure}

Figure \ref{Diagrammatic representation of the current operator.}
depicts the diagrammatic representation of this operator in Fourier
space.

The complexity of this expansion becomes clear. In eq. \ref{current in terms of G<},
both the current operator and the Green's functions follow a perturbation
expansion. Furthermore, each interaction operator in every one of
the terms in the Green's function expansion also follows a similar
expansion. We now have all the objects needed for the perturbative
expansion of the conductivity.

\subsection{Perturbative expansion of the conductivity}

\begin{figure}
\includegraphics[scale=0.53]{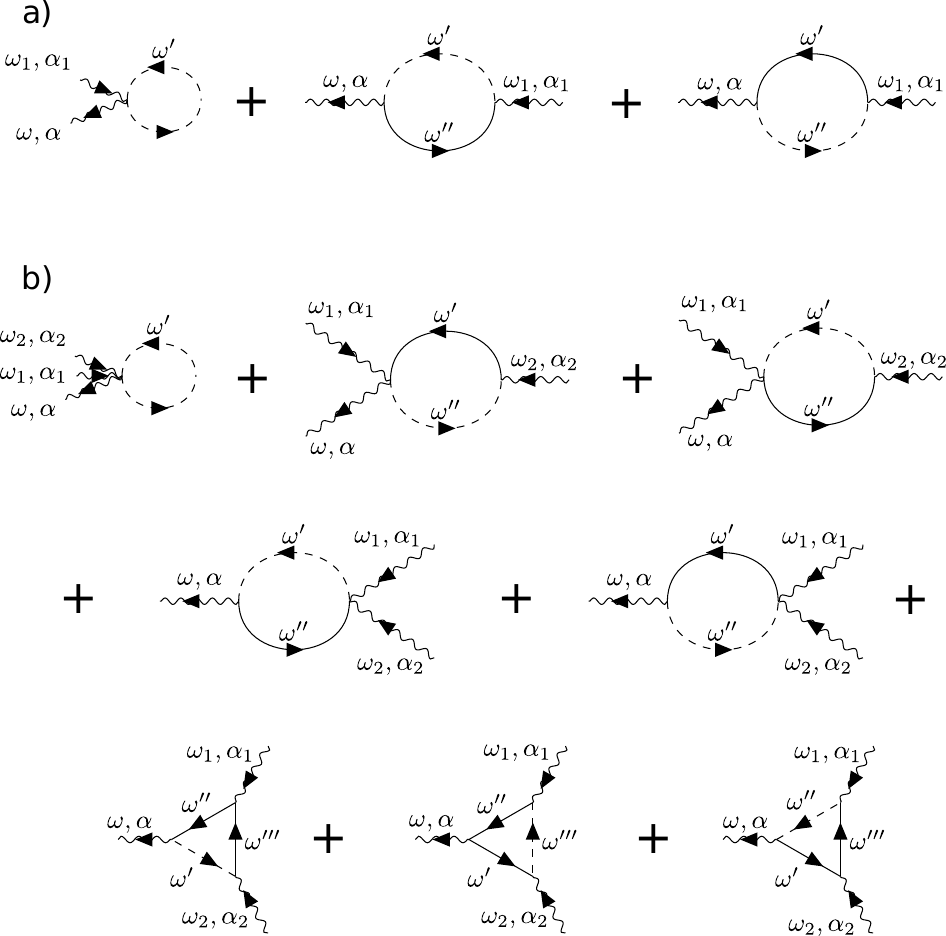}\caption{Expansion of the expected value of the conductivity in (a) first and
(b) second order.}
\label{full expansion}
\end{figure}

In the previous sections we laid out the expressions for each individual
operator in our expansion and represented their corresponding diagrammatic
depictions. In this subsection, we put together all the elements of
the previous sections to provide the full diagrammatic representation
of the first and second-order conductivities. This expansion closely
resembles that of \citep{parkerDiagrammaticApproachNonlinear2018}
but has several differences due to the usage of these specific Green's
functions. The only thing left to do is to replace the perturbed objects
in the diagrammatic representation of the expected value of the current
operator by their expansions. It is straightforward to see how the
diagrams fit together in Figure \ref{full expansion}, which shows
all the contributing diagrams up to second order.

Obtaining the conductivity from the current is a matter of expressing
the frequencies $\omega'$, $\omega''$ and $\omega'''$ in terms
of $\omega_{1}$, $\omega_{2}$ and $\omega$ and using $\boldsymbol{E}\left(\omega\right)=i\omega\boldsymbol{A}\left(\omega\right)$.
The Dirac delta in eq. \ref{conductivity definition} simply means
that $\omega$ is to be replaced by the sum of external frequencies
entering the diagram. Thus, the n-th order conductivity may be found
using the following rules:
\begin{enumerate}
\item Draw all the diagrams with $n$ wavy lines coming in the diagram,
one going out and one dashed interconnecting line. Integrate over
the internal frequencies and ignore the conservation of momentum in
the vertex containing $\omega$, as that is already taken into account
by the Dirac delta in the definition of the conductivity.
\item Reading clockwise starting from the vertex containing $\omega$, insert,
by order, a generalized velocity operator $h^{\alpha_{1}\cdots\alpha_{k}}$
at each vertex and a Green's function at each edge. Each $\alpha_{i}$
is the label of a frequency line connecting to the vertex. If the
edge is a dashed line, the Green's function is $ig^{<}$. All the
edges before that correspond to $ig^{R}$ and the ones after it to
$ig^{A}$. Trace over the resulting operator.
\item Multiply by $\Omega^{-1}e^{n+1}\prod_{k=1}^{n}\left(i\omega_{k}\right)^{-1}\left(i\hbar\right)^{1-N}$,
where $n$ is the number of dashed lines and $N$ is the number of
interconnecting lines. For each vertex, divide by the factorial of
the number of outgoing lines.
\end{enumerate}
Following these rules and replacing $ig^{<}$ by eq. \ref{eq: lesser green's function},
the first-order conductivity is found:

\begin{widetext}
\begin{equation}
\sigma^{\alpha\beta}\left(\omega\right)=\frac{ie^{2}}{\Omega\omega}\int_{-\infty}^{\infty}d\epsilon f(\epsilon)\text{Tr}\left[\hat{h}^{\alpha\beta}\delta\left(\epsilon-H_{0}\right)+\frac{1}{\hbar}\hat{h}^{\alpha}g^{R}\left(\epsilon/\hbar+\omega\right)\hat{h}^{\beta}\delta\left(\epsilon-H_{0}\right)+\frac{1}{\hbar}\hat{h}^{\alpha}\delta\left(\epsilon-H_{0}\right)\hat{h}^{\beta}g^{A}\left(\epsilon/\hbar-\omega\right)\right].\label{cond 1st order}
\end{equation}

\end{widetext}

Similarly, for the second-order conductivity:

\begin{widetext}

\begin{eqnarray}
 &  & \sigma^{\alpha\beta\gamma}\left(\omega_{1},\omega_{2}\right)=\frac{1}{\Omega}\frac{e^{3}}{\omega_{1}\omega_{2}}\int_{-\infty}^{\infty}d\epsilon f(\epsilon)\text{Tr}\left[\frac{1}{2}\hat{h}^{\alpha\beta\gamma}\delta(\epsilon-H)+\frac{1}{\hbar}\hat{h}^{\alpha\beta}g^{R}\left(\epsilon/\hbar+\omega_{2}\right)\hat{h}^{\gamma}\delta(\epsilon-H)\right.\label{eq: cond 2nd order}\\
 &  & +\frac{1}{\hbar}\hat{h}^{\alpha\beta}\delta(\epsilon-H)\hat{h}^{\gamma}g^{A}\left(\epsilon/\hbar-\omega_{2}\right)+\frac{1}{2\hbar}\hat{h}^{\alpha}g^{R}\left(\epsilon/\hbar+\omega_{1}+\omega_{2}\right)\hat{h}^{\beta\gamma}\delta(\epsilon-H)+\frac{1}{2\hbar}\hat{h}^{\alpha}\delta(\epsilon-H)\hat{h}^{\beta\gamma}g^{A}\left(\epsilon/\hbar-\omega_{1}-\omega_{2}\right)\nonumber \\
 &  & +\frac{1}{\hbar^{2}}\hat{h}^{\alpha}g^{R}\left(\epsilon/\hbar+\omega_{1}+\omega_{2}\right)\hat{h}^{\beta}g^{R}\left(\epsilon/\hbar+\omega_{2}\right)\hat{h}^{\gamma}\delta(\epsilon-H)+\frac{1}{\hbar^{2}}\hat{h}^{\alpha}g^{R}\left(\epsilon/\hbar+\omega_{1}\right)\hat{h}^{\beta}\delta(\epsilon-H)\hat{h}^{\gamma}g^{A}\left(\epsilon/\hbar-\omega_{2}\right)\nonumber \\
 &  & \left.+\frac{1}{\hbar^{2}}\hat{h}^{\alpha}\delta(\epsilon-H)\hat{h}^{\beta}g^{A}\left(\epsilon/\hbar-\omega_{1}\right)\hat{h}^{\gamma}g^{A}\left(\epsilon/\hbar-\omega_{1}-\omega_{2}\right)\right].\nonumber 
\end{eqnarray}

\end{widetext}

The procedure is exactly the same for the $n$-th order conductivity,
which will have $2^{n-1}\left(n+2\right)$ diagrams. The higher-order
expansions will not be obtained because a realistic computation of
physical quantities with those formulas using spectral methods would
require tremendous computational power. This point will be further
explained in the next section. Despite these limitations, we have
provided the required framework for obtaining those formulas, which
might be useful in the future.

\section{Spectral methods\label{Section IV}}

From the previous section, it becomes clear that the only objects
needed to calculate the conductivity up to any order are the retarded
and advanced Green's functions, Dirac deltas and the generalized velocity
operators. The meaning of these functions is obvious in the energy
eigenbasis of $H_{0}$, but not in the position basis that we ultimately
want to use. Therefore, these objects are expanded in a truncated
series of Chebyshev polynomials. This choice of polynomials allows
for a very efficient and numerically stable method of computing these
functions \citep{weisse2006kernel}. The fact that the Dirac deltas
and Green's functions have singularities means that their expansions
will be plagued by Gibbs oscillations. Some methods like KPM add a
weight function \citep{gautschi1970construction,sack1971algorithm}
to each term in the expansion, effectively damping the oscillations.
Let's take as an example the Green's function. Using a kernel, as
more polynomials are added to the expansion, the expansion becomes
closer to the exact Green's function, and so more singular. Although
the behavior approaches that of a Green's function as more polynomials
are added, we cannot speak of convergence in the usual sense. In order
to assess the convergence properties of our method, we will not use
a kernel, but instead use a finite imaginary broadening parameter
inside the Green's function, that is $i0^{+}\rightarrow i\lambda$.
For a finite $\lambda$, the function is no longer singular and so
we can expect the expansion to converge within a given accuracy after
enough polynomials have been added. In this paper, we will use an
exact decomposition of the Green's function \citep{PhysRevLett.115.106601}
in terms of Chebyshev polynomials in order to be able to evaluate
the convergence of our method. The term $\hbar/\lambda$ may also
be interpreted as a phenomenological relaxation time due to inelastic
scattering processes and therefore may be adjusted to reflect this
fact.

In this section, we introduce the Chebyshev polynomials and use them
to expand the Green's functions and Dirac deltas of the previous formulas.
This will cast those formulas into a more useful form.

\subsection{Expansion in Chebyshev polynomials}

The Chebyshev polynomials of the first kind are a set of orthogonal
polynomials defined in the range $\left[-1,1\right]$ by

\begin{equation}
T_{n}\left(x\right)=\cos\left(n\arccos\left(x\right)\right).
\end{equation}
They satisfy a recursion relation

\begin{eqnarray}
T_{0}\left(x\right) & = & 1\\
T_{1}\left(x\right) & = & x\\
T_{n+1}\left(x\right) & = & 2xT_{n}\left(x\right)-T_{n-1}\left(x\right)
\end{eqnarray}
and the following orthogonality relation

\begin{equation}
\int_{-1}^{1}T_{n}\left(x\right)T_{m}\left(x\right)\frac{\text{d}x}{\sqrt{1-x^{2}}}=\delta_{nm}\frac{1+\delta_{n0}}{2}.
\end{equation}

These polynomials may be used to expand functions of the Hamiltonian
provided its spectrum has been scaled by a factor $\Delta_{E}$ to
fit in the range $\left[-1,1\right]$. The only functions of the Hamiltonian
appearing in the expansion are Dirac deltas and Green's functions,
which have the following expansions in terms of Chebyshev polynomials:

\begin{equation}
\delta(\epsilon-H_{0})=\sum_{n=0}^{\infty}\Delta_{n}(\epsilon)\frac{T_{n}(H_{0})}{1+\delta_{n0}}\label{eq: delta function cheb expansion}
\end{equation}

\begin{equation}
g^{\sigma,\lambda}(\epsilon,H_{0})=\frac{\hbar}{\epsilon-H_{0}+i\sigma\lambda}=\hbar\sum_{n=0}^{\infty}g_{n}^{\sigma,\lambda}(\epsilon)\frac{T_{n}(H_{0})}{1+\delta_{n0}}\label{eq: green function cheb expansion}
\end{equation}
where

\begin{equation}
\Delta_{n}(\epsilon)=\frac{2T_{n}(\epsilon)}{\pi\sqrt{1-\epsilon^{2}}}
\end{equation}
and

\begin{equation}
g_{n}^{\sigma,\lambda}\left(\epsilon\right)=-2\sigma i\frac{e^{-ni\sigma\arccos\left(\epsilon+i\sigma\lambda\right)}}{\sqrt{1-\left(\epsilon+i\sigma\lambda\right)^{2}}}.
\end{equation}

The function $g^{\sigma,\lambda}$ encompasses both retarded and advanced
Green's functions in the limit $\lambda\rightarrow0^{+}$: $g^{+,0^{+}}$
is the retarded Green's function and $g^{-,0^{+}}$ the advanced one.
The operator part has been completely separated from its other arguments.
All the Dirac deltas and Green's functions may therefore be separated
into a polynomial of $H_{0}$ and a coefficient which encapsulates
the frequency and energy parameters. The trace in the conductivity
now becomes a trace over a product of polynomials and $\hat{h}$ operators,
which can be encapsulated in a new object, the $\Gamma$ matrix:

\begin{equation}
\Gamma_{n_{1}\cdots n_{m}}^{\boldsymbol{\alpha}_{1},\cdots,\boldsymbol{\alpha}_{m}}=\frac{\text{Tr}}{N}\left[\tilde{h}^{\boldsymbol{\alpha}_{1}}\frac{T_{n_{1}}(H_{0})}{1+\delta_{n_{1}0}}\cdots\tilde{h}^{\boldsymbol{\alpha}_{m}}\frac{T_{n_{m}}(H_{0})}{1+\delta_{n_{m}0}}\right].
\end{equation}
The upper indices in bold stand for any number of indices: $\boldsymbol{\alpha}_{1}=\alpha_{1}^{1}\alpha_{1}^{2}\cdots\alpha_{1}^{N_{1}}$.
Here we have used $\tilde{h}^{\boldsymbol{\alpha}_{1}}=\left(i\hbar\right)^{N_{1}}\hat{h}^{\boldsymbol{\alpha}_{1}}$
rather than $\hat{h}$ to avoid using complex numbers when the Hamiltonian
matrix is purely real in our numerical simulations. It's very important
to keep in mind that these new operators are no longer hermitian.
The commas in $\Gamma$ separate the various $\tilde{h}$ operators.
$N$ is the number of unit cells in the sample being studied and ensures
that $\Gamma$ is an intensive quantity. Some examples:

\begin{eqnarray}
\Gamma_{nm}^{\alpha,\beta\gamma} & = & \frac{\text{Tr}}{N}\left[\tilde{h}^{\alpha}\frac{T_{n}(H_{0})}{1+\delta_{n0}}\tilde{h}^{\beta\gamma}\frac{T_{m}(H_{0})}{1+\delta_{m0}}\right]\\
\Gamma_{n}^{\alpha\beta} & = & \frac{\text{Tr}}{N}\left[\tilde{h}^{\alpha\beta}\frac{T_{n}(H_{0})}{1+\delta_{n0}}\right]\\
\Gamma_{nmp}^{\alpha,\beta,\gamma} & = & \frac{\text{Tr}}{N}\left[\tilde{h}^{\alpha}\frac{T_{n}(H_{0})}{1+\delta_{n0}}\tilde{h}^{\beta}\frac{T_{m}(H_{0})}{1+\delta_{m0}}\tilde{h}^{\gamma}\frac{T_{p}(H_{0})}{1+\delta_{p0}}\right].
\end{eqnarray}

The $\Gamma$ matrix only depends on the physical system itself as
it is merely a function of the Hamiltonian and the $\tilde{h}$ operators.
The coefficients of the Chebyshev expansion may similarly be aggregated
into a matrix, which we denote by $\Lambda$. Some examples:

\begin{eqnarray}
\Lambda_{n} & = & \int_{-\infty}^{\infty}d\epsilon f(\epsilon)\Delta_{n}(\epsilon)\\
\Lambda_{nm}\left(\omega\right) & = & \hbar\int_{-\infty}^{\infty}d\epsilon f(\epsilon)\left[g_{n}^{R}\left(\epsilon/\hbar+\omega\right)\Delta_{m}\left(\epsilon\right)\right.\nonumber \\
 &  & \left.\quad\quad\quad\quad+\Delta_{n}\left(\epsilon\right)g_{m}^{A}\left(\epsilon/\hbar-\omega\right)\right]
\end{eqnarray}
and

\begin{eqnarray}
 &  & \Lambda_{nmp}\left(\omega_{1},\omega_{2}\right)=\\
 &  & \hbar^{2}\int_{-\infty}^{\infty}d\epsilon f(\epsilon)\left[g_{n}^{R}\left(\epsilon/\hbar+\omega_{1}+\omega_{2}\right)g_{m}^{R}\left(\epsilon/\hbar+\omega_{2}\right)\Delta_{p}\left(\epsilon\right)\right.\nonumber \\
 &  & \quad\quad\quad\,\,\,\,\,\quad+g_{n}^{R}\left(\epsilon/\hbar+\omega_{1}\right)\Delta_{m}\left(\epsilon\right)g_{p}^{A}\left(\epsilon/\hbar-\omega_{2}\right)\nonumber \\
 &  & \left.\quad\quad\quad\,\,\,\,\,\quad+\Delta_{n}\left(\epsilon\right)g_{m}^{A}\left(\epsilon/\hbar-\omega_{1}\right)g_{p}^{A}\left(\epsilon/\hbar-\omega_{1}-\omega_{2}\right)\right]\nonumber 
\end{eqnarray}

In terms of these new objects, the conductivities become

\begin{equation}
\sigma^{\alpha\beta}\left(\omega\right)=\frac{-ie^{2}}{\Omega_{c}\hbar^{2}\omega}\left[\sum_{n}\Gamma_{n}^{\alpha\beta}\Lambda_{n}+\sum_{nm}\Lambda_{nm}\left(\omega\right)\Gamma_{nm}^{\alpha,\beta}\right]\label{eq:conductivity_1st_order}
\end{equation}
in first order and

\begin{eqnarray}
 &  & \sigma^{\alpha\beta\gamma}\left(\omega_{1},\omega_{2}\right)=\frac{ie^{3}}{\Omega_{c}\omega_{1}\omega_{2}\hbar^{3}}\left[\frac{1}{2}\sum_{n}\Lambda_{n}\Gamma_{n}^{\alpha\beta\gamma}\right.\nonumber \\
 &  & +\sum_{nm}\Lambda_{nm}\left(\omega_{2}\right)\Gamma_{nm}^{\alpha\beta,\gamma}+\frac{1}{2}\sum_{nm}\Lambda_{nm}\left(\omega_{1}+\omega_{2}\right)\Gamma_{nm}^{\alpha,\beta\gamma}\nonumber \\
 &  & \left.+\sum_{nmp}\Lambda_{nmp}\left(\omega_{1},\omega_{2}\right)\Gamma_{nmp}^{\alpha,\beta,\gamma}\right]
\end{eqnarray}
in second order. $\Omega_{c}$ is the volume of the unit cell.

\subsection{Considerations on the numerical storage of $\Gamma$ }

Naturally, one cannot expect to sum the entire Chebyshev series, so
it has to be truncated at a certain number of polynomials $N_{\text{max}}$.
Each of the entries in a $\Gamma$ matrix represents a complex number.
Numerically, this is represented as two double-precision floating-point
numbers, each taking up $8$ bytes of storage. The amount of storage
needed to store a $\Gamma$ matrix of dimension $n$ is $16$$N_{\text{max}}^{n}$.
The number of Chebyshev polynomials needed to obtain a decent resolution
depends heavily on the problem at hand, but a typical number may be
$N_{\text{max}}=1024$. A one-dimensional $\Gamma$ matrix would take
up $16$ KiB of storage, a two-dimensional matrix $16$ MiB and a
three-dimensional matrix $16$ GiB. Three-dimensional matrices appear
in the second-order conductivity. The third-order conductivity would
require a four-dimensional matrix and as such, $16$ TiB of storage.
Numbers like these make it unrealistic to go beyond second order conductivity.

\section{Numerical results\label{Section V}}

In this section we showcase several examples, of increasing complexity,
to compare our formalism with the literature. Starting with graphene,
we compute the linear optical conductivity and verify that it agrees
perfectly with the $\boldsymbol{k}$-space formalism. Breaking the
sublattice symmetry with  gapped graphene, we are able to obtain the
second-order conductivity and check that it too agrees perfectly.
This proves that our method is able to accurately reproduce the existing
results. Then, we show two examples that cannot be reproduced easily
with the $\boldsymbol{k}$-space formalism: second harmonic generation
in gapped graphene with Anderson disorder and vacancies of varying
concentration. Finally, the convergence properties are evaluated and
the efficiency of the method is discussed.

\subsection{Linear optical response in graphene}

\begin{figure}
\begin{centering}
\includegraphics[scale=0.3]{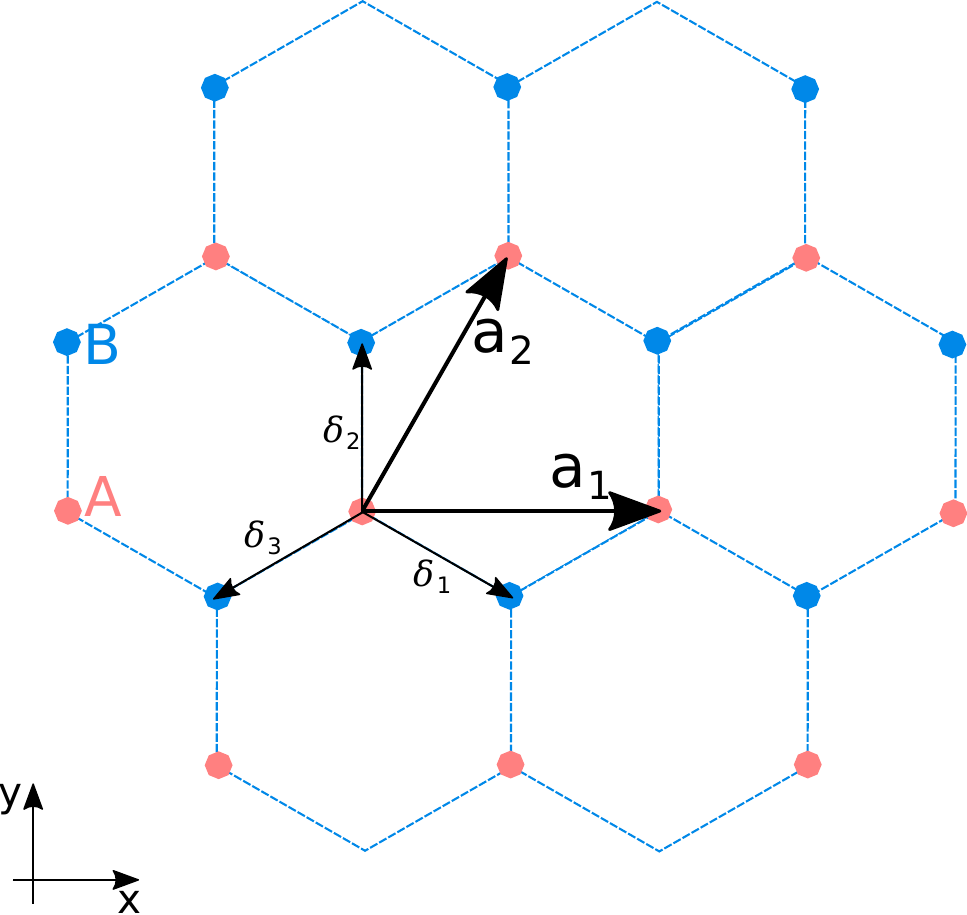}
\par\end{centering}
\caption{Honeycomb lattice and choice of primitive vectors.}
\label{lattice_graphene}\label{dhbn}
\end{figure}

Let $a$ be the distance between consecutive atoms in the honeycomb
lattice. Then, the primitive vectors between unit cells are (see Figure
\ref{lattice_graphene})

\begin{eqnarray*}
\boldsymbol{a}_{1} & = & a\left(\sqrt{3},0\right)\\
\boldsymbol{a}_{2} & = & a\left(\frac{\sqrt{3}}{2},\frac{3}{2}\right)
\end{eqnarray*}
and the distance vectors between nearest neighbors are

\begin{eqnarray*}
\boldsymbol{\delta}_{1} & = & \frac{a}{2}\left(\sqrt{3},-1\right)\\
\boldsymbol{\delta}_{2} & = & a\left(0,1\right)\\
\boldsymbol{\delta}_{3} & = & \frac{a}{2}\left(-\sqrt{3},-1\right).
\end{eqnarray*}

The area of the unit cell is $\Omega_{c}=\frac{3\sqrt{3}}{2}a^{2}$.
Starting from eq. \ref{eq:TB hamiltonian}, the graphene Hamiltonian
is obtained by invoking translational invariance of the unit cell
$t_{\mu\nu}\left(\boldsymbol{R}_{m},\boldsymbol{R}_{n}\right)=t_{\mu\nu}\left(\boldsymbol{R}_{m}-\boldsymbol{R}_{n}\right)$
and

\[
t_{AB}\left(\boldsymbol{\delta}_{1}\right)=t_{AB}\left(\boldsymbol{\delta}_{2}\right)=t_{AB}\left(\boldsymbol{\delta}_{3}\right)=-t.
\]

The remaining non-zero hopping integrals are found by using $t_{AB}=t_{BA}$.
The on-site energies $t_{AA}\left(\boldsymbol{0}\right)$ and $t_{BB}\left(\boldsymbol{0}\right)$
are taken to be zero. A factor of two is included due to spin degeneracy. 

\begin{figure}
\includegraphics[scale=0.3]{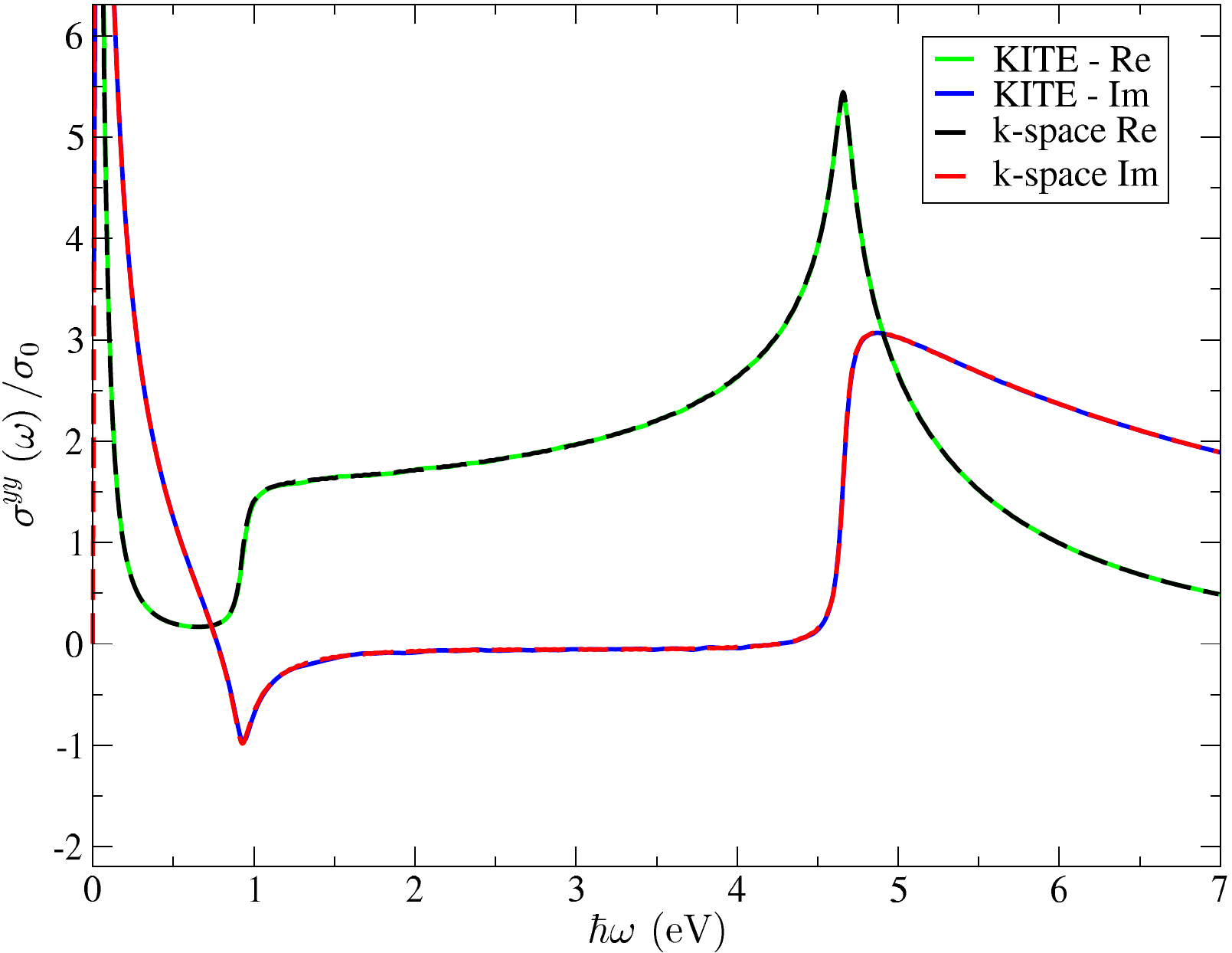}

\caption{First-order longitudinal $yy$ conductivity for graphene in units
of $\sigma_{0}=e^{2}/\hbar$. Hopping parameter: $t=2.33\text{eV}$,
temperature: $T=2.33\text{mK}$, chemical potential: $\mu=0.466\text{eV}$,
broadening parameter: $\lambda=38.8\text{meV}$, number of Chebyshev
moments used: $M=2048$, lattice size: $L=4096\times4096$. The solid
curves represent the optical conductivity obtained by KITE (real part
in green, imaginary in blue). The superimposed dashed lines are obtained
in \citep{PhysRevB.96.035431}.}
\label{graphene_conductivity_image}
\end{figure}

These parameters were used to obtain the first-order optical conductivity
for graphene, as seen in Figure \ref{graphene_conductivity_image}.
We used a lattice with $4096$ unit cells in each direction and $2048$
Chebyshev moments in the expansion. The resulting plot is compared
to the results obtained in \citep{PhysRevB.96.035431} through $\boldsymbol{k}$-space
integration of a translation-invariant system. The curves are indistinguishable.

\subsection{Gapped graphene}

The only difference relative to regular graphene is found in the on-site
energies. Let $t_{AA}\left(\boldsymbol{0}\right)=\Delta/2$ and $t_{BB}\left(\boldsymbol{0}\right)=-\Delta/2$.
With the opening of a sizeable gap, the effects of excitons become
relevant. In this simplified model, we ignore these effects to focus
on the intrinsic second-order respose of the material. In these conditions,
the one- and three-dimensional $\Gamma$ matrices are identically
zero\footnote{These matrices were explicitly calculated in the k basis and shown
to be exactly zero.}, so the second-order conductivity may be calculated resorting only
to two-dimensional $\Gamma$ matrices. The calculation is thus simplified
tremendously because the second-order conductivity reduces to
\begin{eqnarray*}
 &  & \sigma^{\alpha\beta\gamma}\left(\omega_{1},\omega_{2}\right)=\frac{ie^{3}}{\Omega_{c}\omega_{1}\omega_{2}\hbar^{3}}\times\\
 &  & \left[\sum_{nm}\Lambda_{nm}\left(\omega_{2}\right)\Gamma_{nm}^{\alpha\beta,\gamma}+\frac{1}{2}\sum_{nm}\Lambda_{nm}\left(\omega_{1}+\omega_{2}\right)\Gamma_{nm}^{\alpha,\beta\gamma}\right].
\end{eqnarray*}

\begin{figure}
\includegraphics[scale=0.3]{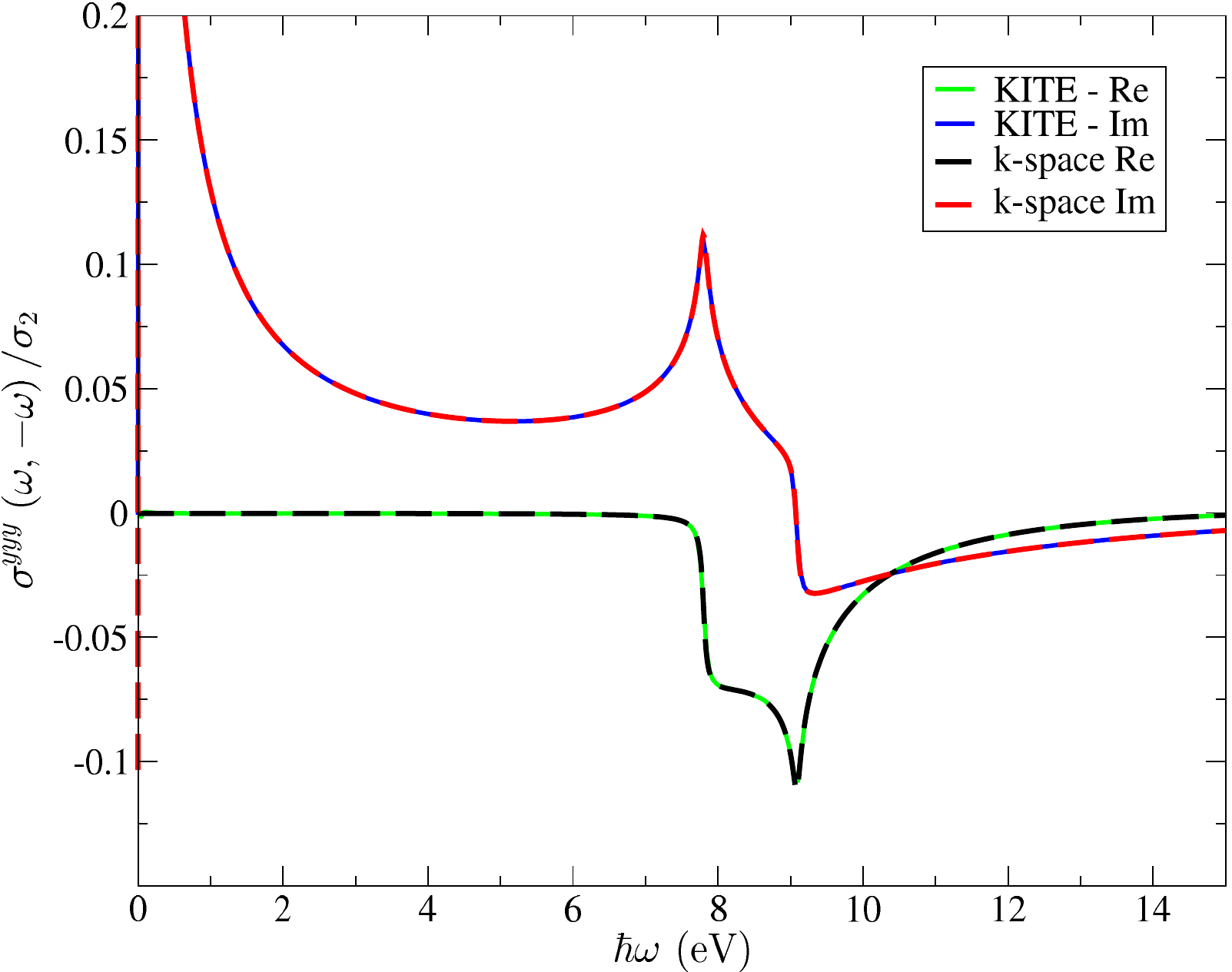}\caption{Second-order $yyy$ photogalvanic effect for gapped graphene. Hopping
parameter: $t=2.33\text{eV}$, temperature: $T=0\text{K}$, chemical
potential: $\mu=0\text{eV}$, gap $\Delta=7.80\text{eV}$ broadening
parameter: $\lambda=39\text{meV}$, number of Chebyshev moments used:
$M=2048$, lattice size: $L=4096\times4096$. The imaginary part disappears
after the result is properly symmetrized.}

\label{h-BN_photoconductivity_image}
\end{figure}

The indices $n,m$ are understood to be summed over. The photogalvanic
effect may be reproduced from this formula by setting $\omega_{1}=\omega=-\omega_{2}$
and the numerical results are shown in Figure \ref{h-BN_photoconductivity_image}.
Again, we used $4096$ unit cells in each lattice direction and $2048$
Chebyshev moments and compare the results with the ones obtained by
integrating in $\boldsymbol{k}$-space, just like in the previous
subsection. For convenience, we define the constant $\sigma_{2}=e^{3}a/4t\hbar$
\citep{hipolito2016nonlinear} in terms of the hopping integral $t$
and the lattice parameter $a$.

 This particular example benefits considerably from the cancellation
of the most complicated objects that needed to be calculated. In appendix
\ref{SEC Appendix-A} we present an example with less symmetry that
confirms the complete agreement between our method and the $\boldsymbol{k}$-space
formalism.

\subsection{Photogalvanic effect in gapped graphene with Anderson disorder}

\begin{figure}
\includegraphics[scale=0.3]{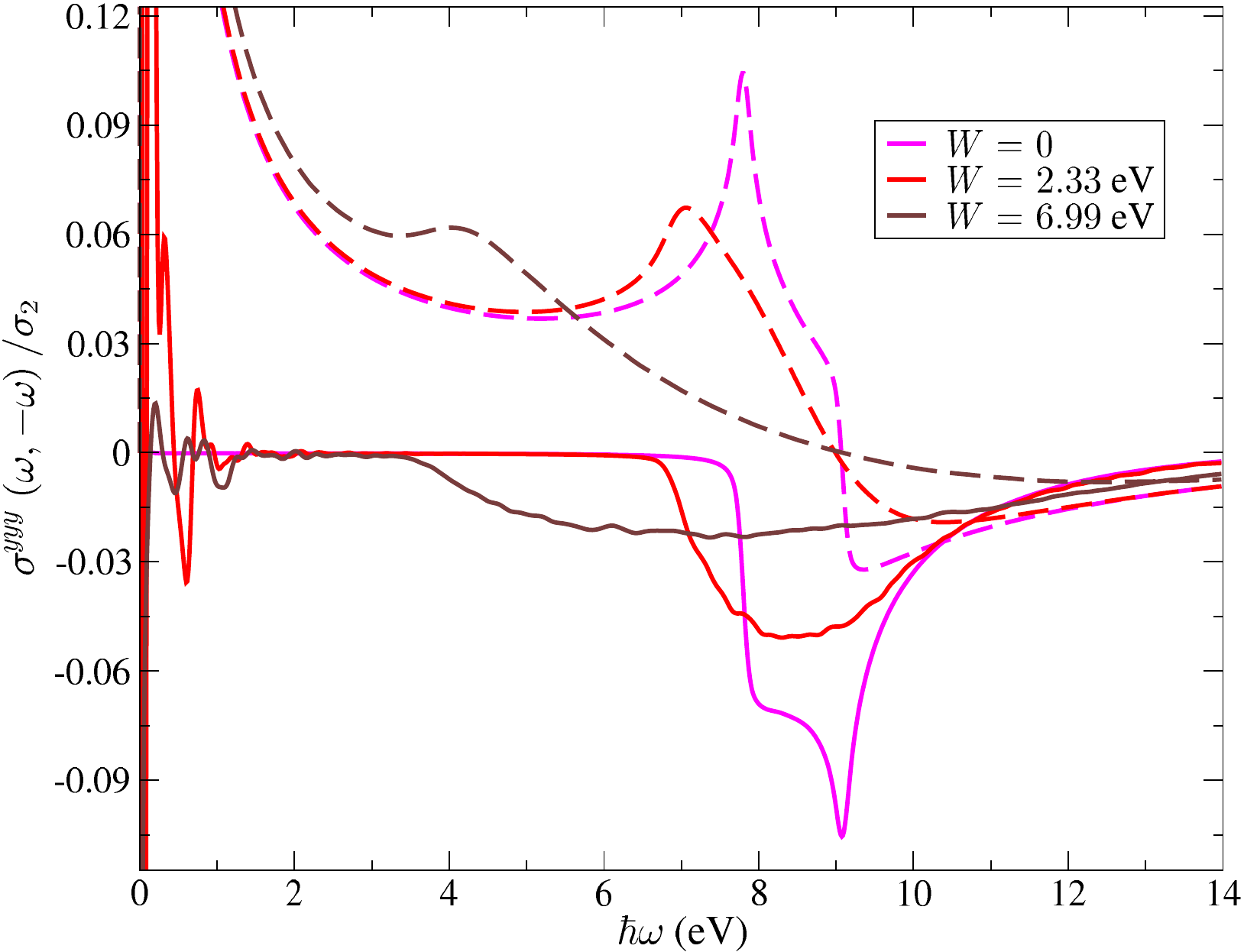}\caption{Photogalvanic effect for gapped graphene in the presence of Anderson
disorder of varying strength $W$ and a broadening parameter of $\lambda=23\text{meV}$.
The parameters are the same as for Figure \ref{h-BN_photoconductivity_image}
except for the number of polynomials, which is $M=512$. The dashed
lines represent the imaginary part of the conductivity.}
\label{Anderson disorder}
\end{figure}

Our formalism does not rely on translation invariance, and so may
be used to study disordered systems. To show this, we now introduce
to gapped graphene a simple model for disorder by letting each atomic
site have a random local energy taken from a uniform distribution
$\left[-W/2,W/2\right]$ (Anderson disorder \citep{PhysRev.109.1492}):

\[
\mathcal{H}_{W}=\sum_{\boldsymbol{R}}\sum_{\sigma}W_{\sigma}\left(\boldsymbol{R}\right)c_{\sigma}^{\dagger}\left(\boldsymbol{R}\right)c_{\sigma}\left(\boldsymbol{R}\right)
\]
where $\boldsymbol{R}$ is the position of the unit cell and $\sigma$
labels the atoms inside each unit cell. The presence of disorder is
expected to smooth out the sharp features of the  optical response.
As disorder increases, we should see a decrease in conductivity due
to Anderson localization. This is the exact behavior that is seen
in Figure \ref{Anderson disorder} where we plot the photogalvanic
effect in gapped graphene in the presence of Anderson disorder of
varying strength. Some fluctuations exist at the features, which are
expected to disappear as the system size approaches the thermodynamic
limit.

As expected, the introduction of Anderson disorder produces a broadening
of the sharp features of the nonlinear optical conductivity. This
broadening also means that there will be a larger response to the
external electric field at frequencies smaller than the gap. 

The large oscillations near the origin reveal something interesting
about the numerical details of our formalism. Eq. \ref{eq: cond 2nd order}
is comprised of a complicated sum of several terms. Individually,
some of these terms may be very large, but there may be cancellations
among them. For each of these terms, the Chebyshev expansion is exact
in the limit of infinite polynomials. For a finite number of polynomials,
there will be slight differences between the exact result and the
expansion, and if the exact result is very large, this difference
will be considerable. It is highly unlikely that this difference will
be the same for each term, and so their sum may not cancel out in
the end. This is the typical behavior at the lower-frequency regime
that is related to the singularities that plagued the velocity gauge
approach. This has been discussed since the early work of Sipe and
challenged for a long time the equivalence between the velocity and
length gauges \citep{PhysRevB.48.11705,PhysRevB.96.035431,passosNonlinearOpticalResponses2018}.
This effect could be fully mitigated by greatly increasing the number
of polynomials, but here we are interested in the finite frequency
behavior.

\subsection{Second-harmonic generation of gapped graphene with vacancies}

\begin{figure}
\includegraphics[scale=0.3]{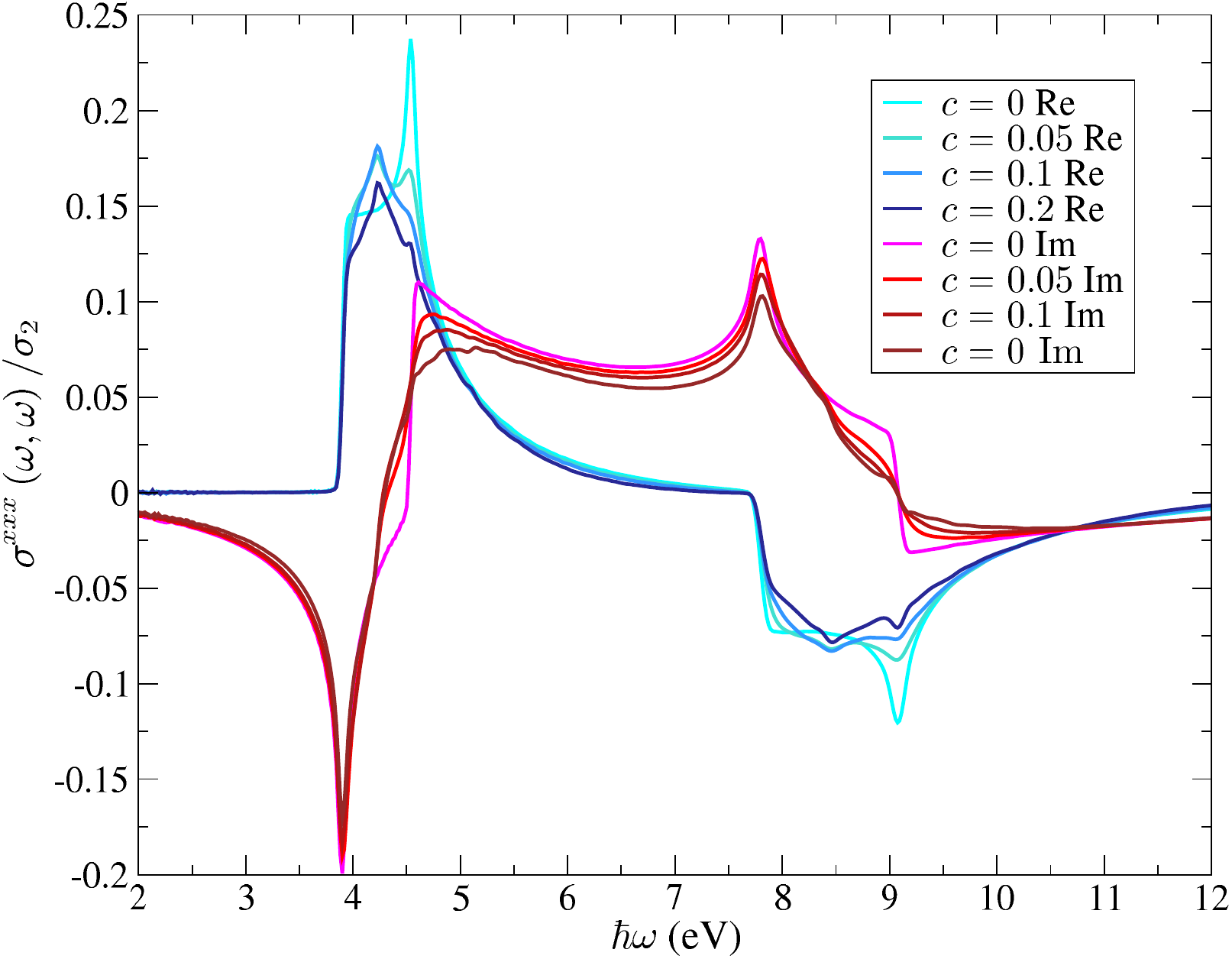}

\caption{Second-harmonic generation in  gapped graphene for a varying concentration
of vacancies and $\lambda=2.3\text{meV}$. The blue (red) curves represent
the real (imaginary) part of the conductivity. The darker curves have
a higher concentration of vacancies. System size: $L=2048$, number
of polynomials: $M=512$. All the other parameters are the same as
in Figure \ref{Anderson disorder}.}
\label{FIG vacancies}
\end{figure}

In realistic samples, vacancies and impurities may exist due to imperfections
in the fabrication process, as well as other more complex structural
defects. In this section, we show that our method allows us to obtain
the second-harmonic generation of a system with structural disorder.
Using eq. \ref{eq: cond 2nd order}, we show in Figure \ref{FIG vacancies}
the effect of vacancies of varying concentration in the SHG of gapped
graphene. Unlike Anderson disorder, the addition of vacancies to the
system does not change the gap. Their most noticeable effect is to
flatten the features of the second-harmonic generation. As discussed
in the previous section, the lower frequencies are dominated by oscillations
and would require many more polynomials to fully converge. Therefore,
we omit that region and only represent the remaining regions, which
have already converged within the desired accuracy.

\subsection{Considerations on convergence and accuracy}

\begin{figure}
\includegraphics[scale=0.3]{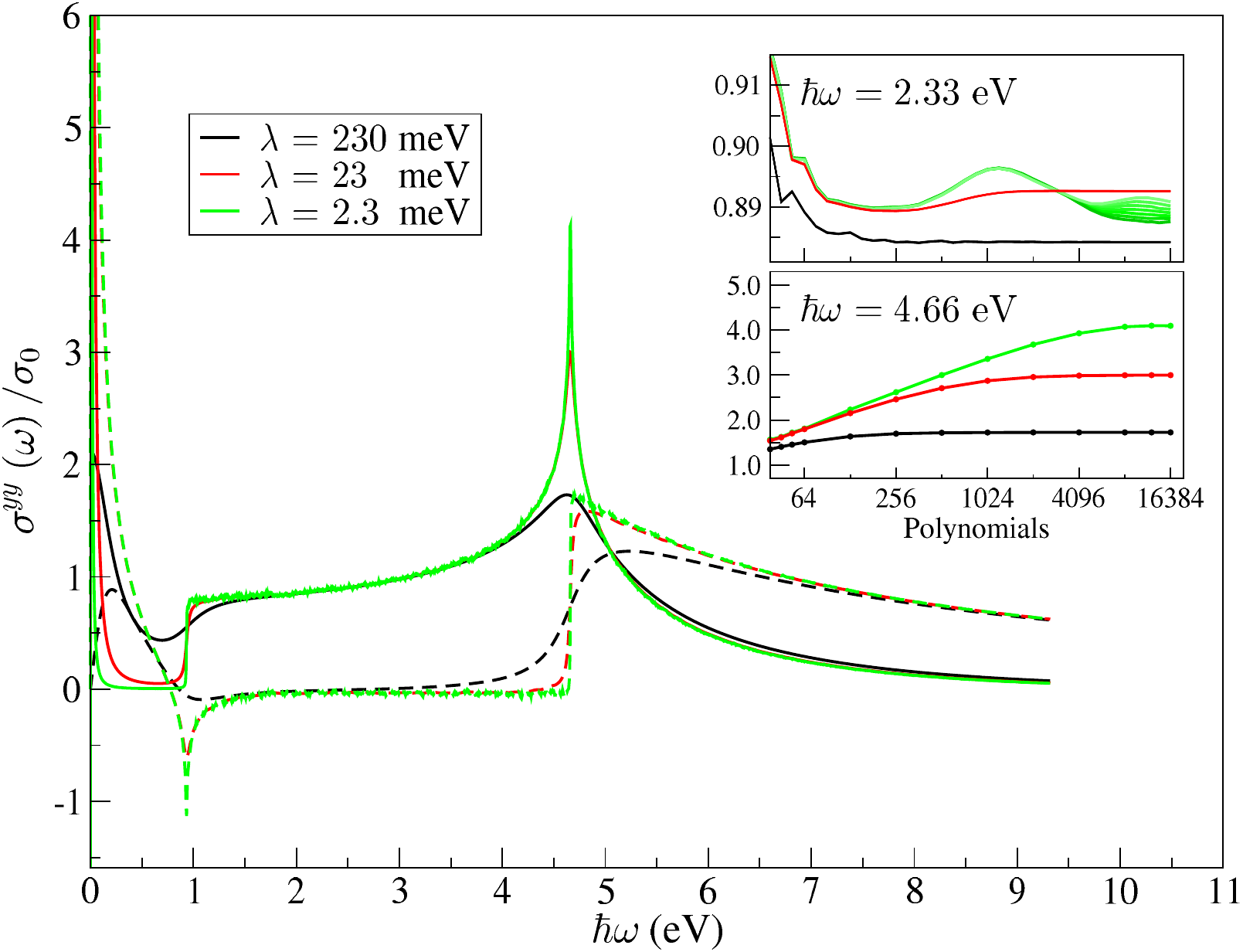}

\caption{First-order optical $yy$ conductivity per spin of graphene for $M=16384$
and $L=2048$ now as a function of the broadening parameter $\lambda$.
The remaining parameters remain the same as for Figure \ref{graphene_conductivity_image}.
The solid (dashed) curves represent the real (imaginary) part of the
conductivity. The legend shows the values for the broadening parameter.
The lower inset shows the evolution of the value of the conductivity
for each $\lambda$ as the number of polynomials is increased for
$\hbar\omega=2.33\text{eV}$. The upper inset shows the same thing
but for several very close frequencies around $\hbar\omega=4.66\text{eV}$.
The darker curves correspond to higher frequencies.}
\label{graphene_convergence}
\end{figure}

In this section, we briefly discuss some convergence properties of
our method. For a more thorough discussion, see \citep{weisse2006kernel}.
The convergence to the exact value depends on several factors:
\begin{enumerate}
\item Spectral methods rely on the self-averaging properties of random vectors,
yielding an associated variance. The error bar decreases as $1/\sqrt{N_{R}N}$,
where $N_{R}$ is the number of random vectors and $N$ is the size
of the sample.
\item In the thermodynamic limit of an infinite lattice, the spectrum becomes
continuous and so we expect the conductivity curve to be smooth. However,
the systems used in simulations are finite and so have a typical energy
level spacing, which we denote by $\delta\varepsilon$. This has important
consequences for the resolution. Details characterized by a smaller
energy scale than that of $\delta\varepsilon$ are meaningless because
they cannot be distinguished from the contribution of individual energy
levels. The maximum resolution is therefore limited by the energy
level spacing. For our concrete examples with the honeycomb lattice,
we use $\delta\varepsilon=3\pi t/L$, the energy level spacing at
the Dirac point in graphene for a system of linear dimension $L$.
\item The resolution may be controlled through $\lambda$, the broadening
parameter of the Green's functions. Energy differences smaller than
$\lambda$ become indistinguishable from one another. On the one hand,
a small $\lambda$ is required in order to resolve the sharp features
of the curve accurately. On the other hand, when $\lambda\lesssim\delta\varepsilon$,
the discrete nature of the spectrum starts to become visible through
the roughness of the curve. For sufficiently small $\lambda$, the
expected sharp features of the curve become indistinguishable from
the contributions of the individual energy levels. If these issues
are not solved, they become a major source of systematic error in
the final results. Therefore, if we want to see the expected thermodynamic
limit, we have to ensure $\lambda\gtrsim\delta\varepsilon$.
\end{enumerate}
In Figure \ref{graphene_convergence}, the $yy$ optical conductivity
of graphene is represented for several values of $\lambda$. In this
example, $\delta\varepsilon=5.3\text{meV}$. As $\lambda$ is decreased,
the curve becomes sharper, but when $\lambda=2.3\text{meV}$ the discreteness
of the spectrum starts to become noticeable through the roughness
of the curve. It is starting to diverge from the expected smooth curve
of the thermodynamic limit.

In the lower inset, we study the convergence as a function of the
number of polynomials at $\hbar\omega=4.66\text{eV}$, a region of
rapidly changing conductivity. The smaller the $\lambda$, the more
polynomials are required in order to have a fully converged result.
Within the accuracy $\delta\sigma/\sigma_{0}\simeq0.1$, all the curves
have already converged at $1.6\times10^{4}$ polynomials. These calculations
were repeated for several different initial random vectors. In the
plot we show only one of these calculations. The error bar associated
with the random vectors is too small to be distinguished from the
curves themselves.

In the upper inset, we do the same thing, but now in a very small
region around $\hbar\omega=2.33\text{eV}$, a region of slowly increasing
conductivity. The plot shows three sets of curves with different colors.
Inside each set, we represent a collection of frequencies, ranging
from $\hbar\omega=2.3300\text{eV}$ to $\hbar\omega=2.3316\text{eV}$.
The darker curves correspond to higher frequencies. The main graph
shows that all these curves have converged to the same value in a
region of slowly increasing conductivity. The inset, however, shows
a different picture. The red ($\lambda=23\text{meV}$) and black ($\lambda=230\text{meV}$)
sets of curves show a variation consistent with the expected increasing
conductivity. If one zooms in to those sets of curves, it is possible
to check that they are indeed increasing in value as $\omega$ increases.
The green curve ($\lambda=2.3\text{meV}$) is not only changing in
a scale much larger than expected, but it is also decreasing. This
variation comes from the individual contribution of the energy levels,
not from features of the conductivity and is therefore artificial.
Within the accuracy $\delta\sigma/\sigma_{0}\simeq10^{-3}$ each of
these curves has completely converged at $1.6\times10^{4}$ polynomials
but this level of accuracy is meaningless for $\lambda=2.3\text{meV}$.
The error bars are not shown for clarity, but their values are the
following: at $\lambda=230\text{meV}$, $\delta\sigma/\sigma_{0}=10^{-3}$;
at $\lambda=23\text{meV}$, $\delta\sigma/\sigma_{0}=3\times10^{-3}$;
at $\lambda=2.3\text{meV}$, $\delta\sigma/\sigma_{0}=5\times10^{-3}$.
At this scale, the error bars are comparable to the variation due
to the number of polynomials and to the value of $\lambda$.

These frequencies were chosen to compare the conductivity in a place
where it is expected to converge quickly and another where it is expected
to converge slowly. Looking at these graphs, it is possible to estimate
how many polynomials are required to converge to the final value of
the conductivity for the specified parameter $\lambda$ within a given
accuracy. A rough estimate of the scaling is given by $N\sim\lambda^{-1}$. 

In sum, given a fixed resolution $\lambda$, the number of polynomials
should be large enough to ensure that the curves have converged, and
the system size $L$ should be large enough to ensure that the discreteness
of the spectrum cannot be seen.

A similar analysis may be done for the second-order conductivity.
We will not present it here for two reasons. Firstly, the main points
of the previous paragraphs remain the same. Secondly, we cannot do
such an analysis because the computational cost would be tremendously
higher.

\subsection{Considerations on efficiency}

Our formalism provides a very general framework with which to compute
the nonlinear optical response up to any order. Once the formulas
were obtained, we chose to use spectral methods to perform the computation.
This is not always the most efficient approach: for systems with translation
invariance and periodic boundary conditions, we can specify the formulas
for $\boldsymbol{k}$-space and then perform the explicit integration.
Then, for a given set of parameters (temperature, broadening, Fermi
energy) the computation time will scale as $L^{D}N_{\omega}$ where
$L^{D}$ is the number of points in the Brillouin zone (which is also
the number of lattice sites), $D$ is the dimensionality and $N_{\omega}$
is the number of frequencies we want to compute. For each $\boldsymbol{k}$
and each frequency, this method comes down to diagonalizing the Bloch
Hamiltonian $H_{\boldsymbol{k}}$, and then summing over the whole
set of $\boldsymbol{k}$ points. This method is extremely efficient
at computing the optical conductivity at any order using the velocity
gauge.

Using spectral methods, the computation is split into the calculation
of the Chebyshev moments and the final matrix product of the $\Gamma$
matrices with the $\Lambda$ matrices. The first part is the most
demanding and is independent of the parameters mentioned above. Its
computation time scales as $L^{D}N^{n+1}$, where $n$ is the order
of the conductivity and $N$ is the number of Chebyshev polynomials.
More concretely, if we want to calculate the conductivity for a certain
$\lambda$, using $N\sim\lambda^{-1}\sim N_{\omega}$, we find that
the $\boldsymbol{k}$-space calculation scales much more favorably.

If the system has no translation invariance, $\boldsymbol{k}$-space
integration is no longer useful and we would need to numerically diagonalize
the full Hamiltonian. This method scales as $L^{3D}N_{\omega}$ which
is highly unfavorable and because of that we would be limited to very
small systems. In this context, spectral methods become the preferred
choice.

For the examples used in this paper, the computation of the second-order
conductivity with the $\boldsymbol{k}$-space formalism in a system
with $L=2048$ took around 2 minutes on a Xeon E5-2650 with 16 threads.
In comparison, the same computation took $3$ hours for translation-invariant
gapped graphene with $2048$ polynomials, and $70$ hours for gapped
graphene with Anderson disorder/vacancies and $512$ polynomials.
Despite the discrepancy in computational efficiency, we know of no
other more efficient way to compute the nonlinear optical conductivity
for disordered systems.

\section{Conclusions\label{Section VI}}

We developed an out-of-equilibrium expansion of the non-linear optical
response of non-interacting systems using the Keldysh formalism and
expressed everything in terms of traces of operators. This provides
a basis-independent expression for the linear and non-linear optical
conductivities. This drops the requirement of translation invariance
and allows us to include magnetic fields and disorder in our tight-binding
calculations. We also provide a diagrammatic representation of this
expansion, which makes it a very straightforward process to obtain
those expressions.

The expressions for the non-linear conductivities are calculated numerically
with resort to an expansion in Chebyshev polynomials and a stochastic
evaluation of the trace, in close resemblance to the Kernel Polynomial
Method (KPM). We provide the mapping that takes the aforementioned
expressions and converts them to a numerically-suited object to be
calculated with spectral methods. This is only possible because of
the careful way in which these expressions were constructed in the
first place.

We built an open-source software that is able to calculate the first-
and second-order optical conductivities of very large 2D tight-binding
systems ($10^{10}$ atoms) with disorder and magnetic fields. This
software is used to obtain the first- and second-order conductivities
of graphene and gapped graphene. These same quantities were calculated
independently with the usual integration in $\boldsymbol{k}$-space
and the results are in great agreement, proving the validity of our
method. Finally, we show two examples with disorder that cannot be
treated under the usual $\boldsymbol{k}$-space formalism: gapped
graphene with Anderson disorder and vacancies.

We briefly discuss the convergence properties of this method by analyzing
how the curves change as the resolution is increased. The resolution
is controlled by $\lambda$, the broadening parameter of the Green's
functions and is limited by $\delta\varepsilon$, the energy level
spacing of the system. The expected thermodynamic limit is obtained
by decreasing $\delta\varepsilon$ (increasing the system size) while
ensuring $\delta\varepsilon\lesssim\lambda$.

For systems with translation invariance, the $\boldsymbol{k}$-space
integration is very quick and is preferred over our method. If the
systems do not have this property, this method becomes the most efficient
one to calculate the second-order conductivity with disorder. This
paper serves as a proof of concept and the effects of realistic disorder
on the nonlinear optical properties of 2D materials will be explored
in a future paper.

\section{Acknowledgements}

The work of SMJ is supported by Fundação para a Ciência e Tecnologia
(FCT) under the grant PD/BD/142798/2018. The authors acknowledge financing
of Fundação da Ciência e Tecnologia, of COMPETE 2020 program in FEDER
component (European Union), through projects POCI-01-0145-FEDER-028887
and UID/FIS/04650/2013. The authors also acknowledge financial support
from Fundação para a Ciência e Tecnologia, Portugal, through national
funds, co-financed by COMPETE-FEDER (grant M-ERA-NET2/0002/2016 --
UltraGraf) under the Partnership Agreement PT2020. We would also like
to thank JMB Lopes dos Santos, JPS Pires, GB Ventura, DJ Passos, B
Amorim and D Parker for their helpful comments and suggestions regarding
this paper.

\section{Appendix A}

\subsection{Sublattice displacement\label{SEC Appendix-A}}

\begin{figure}
\includegraphics[scale=0.25]{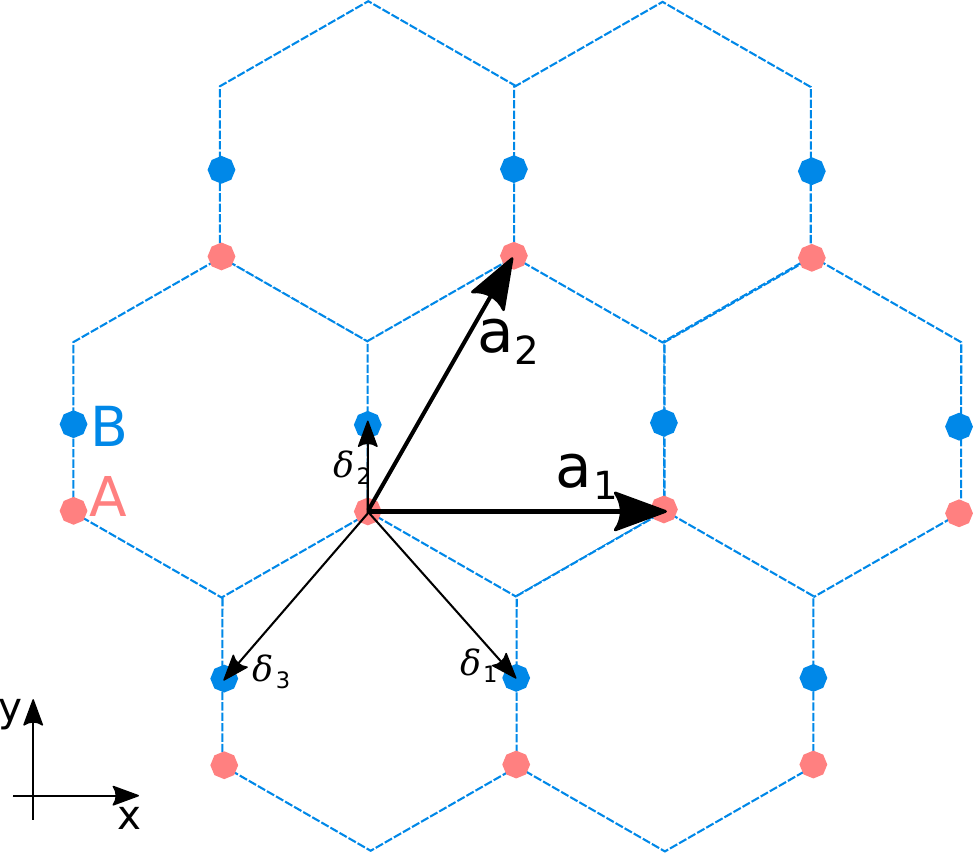}

\caption{Displaced honeycomb lattice and choice of primitive vectors.}

\label{FIG dhbn lattice}
\end{figure}

The calculation of the photogalvanic effect for gapped graphene was
very efficient due to the cancellation of the three-dimensional $\Gamma$
matrices. In this appendix, we provide an extra example, which does
not benefit from that property. By changing the relative position
of the two sublattices, we are able to obtain non-zero values in all
the $\Gamma$ matrices, which enables us to test the remainder of
the formula. All the hopping parameters in this system are exactly
the same as in regular gapped graphene. The only difference is in
the distance between atoms, which changes the velocity operators while
keeping the Hamiltonian intact (See Figure \ref{FIG dhbn lattice}).

The primitive vectors are identical, but the nearest-neighbor vectors
are different:

\begin{eqnarray*}
\boldsymbol{\delta}_{1} & = & a\left(\frac{\sqrt{3}}{2},-1\right)\\
\boldsymbol{\delta}_{2} & = & a\left(0,\frac{1}{2}\right)\\
\boldsymbol{\delta}_{3} & = & a\left(-\frac{\sqrt{3}}{2},-1\right).
\end{eqnarray*}

\begin{figure}
\includegraphics[scale=0.3]{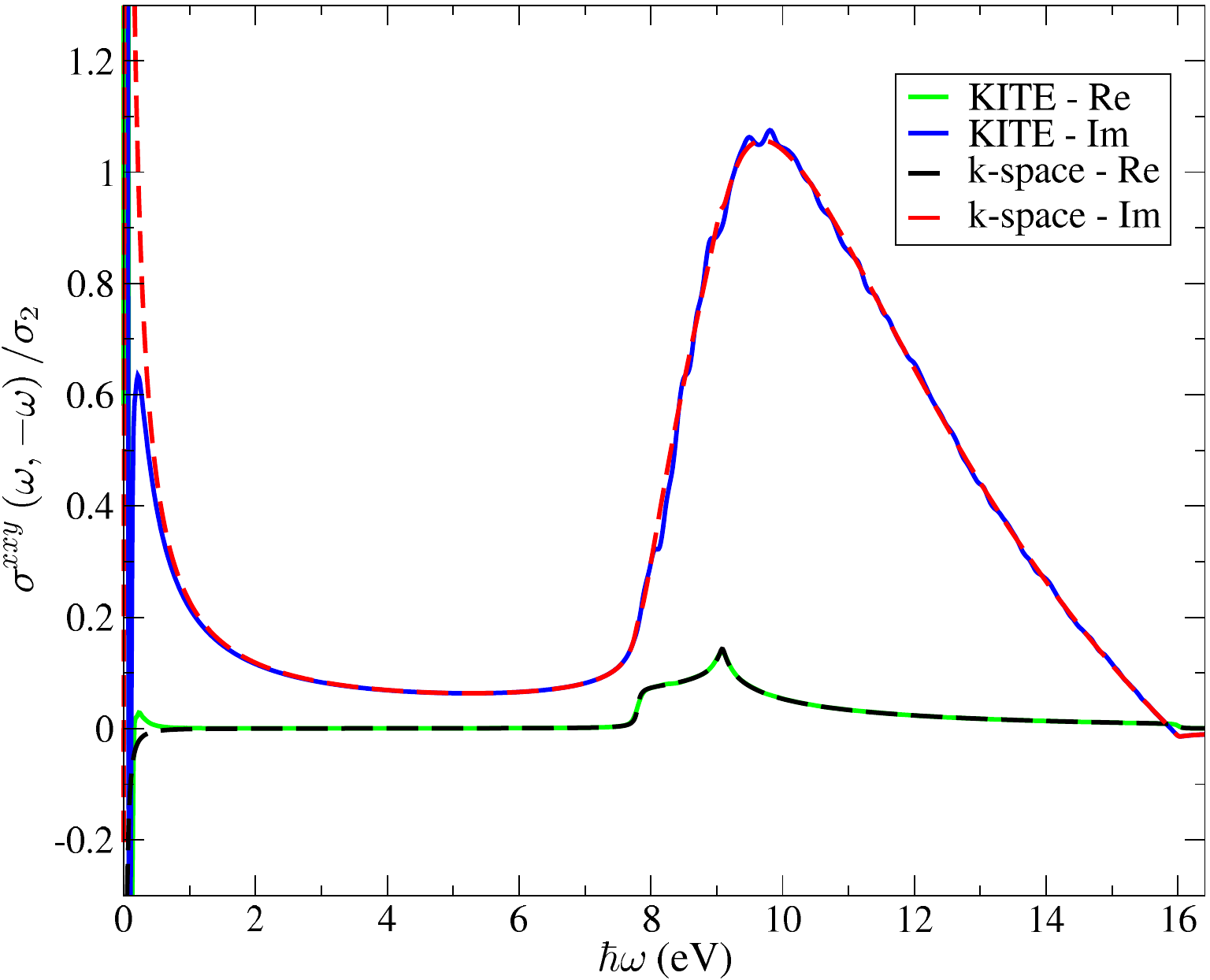}\caption{Second-order $xxy$ photogalvanic effect for displaced gapped graphene.
Hopping parameter: $t=2.33\text{eV}$, temperature: $T=0\text{K}$,
chemical potential: $\mu=0\text{eV}$, gap $\Delta=7.8\text{eV}$
broadening parameter: $\lambda=39\text{meV}$, number of Chebyshev
moments used: $M=512$, lattice size: $L=1024\times1024$.}

\label{dh-BN_photoconductivity_image}
\end{figure}

One of the sublattices was translated in the $y$ direction by $a/2$.
The second-order $xxx$ conductivity remains zero, but now the $xxy$
photogalvanic effect is no longer zero and can be seen in Figure \ref{dh-BN_photoconductivity_image}.
The lattice size and number of polynomials used was reduced to $1024$
and $512$ respectively, due to the greatly increased computational
cost. At lower frequencies, the results start to diverge because there
are not enough polynomials to resolve this region. The results are
in great agreement with the ones obtained by $\boldsymbol{k}$-space
integration. The small oscillations in the imaginary part are expected
to disappear as the number of polynomials is increased.

\bibliographystyle{iopart-num}
\bibliography{bibliography}

\providecommand{\newblock}{}
\begin{thebibliography}{10}
\expandafter\ifx\csname url\endcsname\relax
  \def\url#1{{\tt #1}}\fi
\expandafter\ifx\csname urlprefix\endcsname\relax\def\urlprefix{URL }\fi
\providecommand{\eprint}[2][]{\url{#2}}

\bibitem{PhysRevLett.7.118}
Franken P~A, Hill A~E, Peters C~W and Weinreich G 1961 {\em Phys. Rev. Lett.\/}
  {\bf 7} 118--119

\bibitem{bass1962optical}
Bass M, Franken P, Ward J and Weinreich G 1962 {\em Physical Review Letters\/}
  {\bf 9} 446

\bibitem{new1967optical}
New G and Ward J 1967 {\em Physical Review Letters\/} {\bf 19} 556

\bibitem{PhysRevB.48.11705}
Sipe J~E and Ghahramani E 1993 {\em Phys. Rev. B\/} {\bf 48} 11705--11722

\bibitem{0953-8984-20-38-384204}
Mikhailov S~A and Ziegler K 2008 {\em Journal of Physics: Condensed Matter\/}
  {\bf 20} 384204

\bibitem{GLAZOV2014101}
Glazov M and Ganichev S 2014 {\em Physics Reports\/} {\bf 535} 101--138 ISSN
  0370-1573 high frequency electric field induced nonlinear effects in graphene

\bibitem{bonaccorsoGraphenePhotonicsOptoelectronics2010}
Bonaccorso F, Sun Z, Hasan T and Ferrari A 2010 {\em Nature Photonics\/} {\bf
  4}

\bibitem{PhysRevB.96.035431}
Ventura G~B, Passos D~J, {Lopes dos Santos} J~M~B, Viana Parente~Lopes J~M and
  Peres N~M~R 2017 {\em Phys. Rev. B\/} {\bf 96} 035431

\bibitem{peierls1933theorie}
Peierls R 1933 {\em Zeitschrift f{\"u}r Physik\/} {\bf 80} 763--791

\bibitem{passosNonlinearOpticalResponses2018}
Passos D~J, Ventura G~B, Lopes J~M~V~P, dos Santos J~M~B~L and Peres N~M~R 2018
  {\em Physical Review B\/} {\bf 97}

\bibitem{keldysh1964diagram}
Keldysh L~V 1964 {\em Zh. Eksp. Teor. Fiz.\/} {\bf 47} 1018

\bibitem{weisse2004chebyshev}
Weisse A 2004 {\em The European Physical Journal B-Condensed Matter and Complex
  Systems\/} {\bf 40} 125--128

\bibitem{quantumkite}
Lopes J, Jo{\~a}o S, Ferreira A, Covaci L, Andelkovic M and Rappoport T 2018
  Quantum {{KITE}} https://quantum-kite.com/

\bibitem{jishi2013feynman}
Jishi R~A 2013 {\em Feynman Diagram Techniques in Condensed Matter Physics\/}
  ({Cambridge University Press})

\bibitem{devreeseLinearNonlinearResponse1975}
Devreese J~T and {van Doren} V~E 1975 Linear and nonlinear response theory with
  applications {\em Linear and {{Nonlinear Electron Transport}} in
  {{Solids}}\/} ({\em {{NATO Advanced Study Institute}}, {{Series B}}:
  {{Physics}}\/} vol~17) ({New York: Plenum Press}) pp 3--32 ISBN
  978-1-4757-0877-6

\bibitem{parkerDiagrammaticApproachNonlinear2018}
Parker D~E, Morimoto T, Orenstein J and Moore J~E 2018 {\em Physical Review
  B\/}

\bibitem{weisse2006kernel}
Weisse A, Wellein G, Alvermann A and Fehske H 2006 {\em Reviews of modern
  physics\/} {\bf 78} 275

\bibitem{gautschi1970construction}
Gautschi W 1970 {\em Mathematics of Computation\/} {\bf 24} 245--260

\bibitem{sack1971algorithm}
Sack R and Donovan A 1971 {\em Numerische Mathematik\/} {\bf 18} 465--478

\bibitem{PhysRevLett.115.106601}
Ferreira A and Mucciolo E~R 2015 {\em Phys. Rev. Lett.\/} {\bf 115} 106601

\bibitem{hipolito2016nonlinear}
Hipolito F, Pedersen T~G and Pereira V~M 2016 {\em Physical Review B\/} {\bf
  94} 045434

\bibitem{PhysRev.109.1492}
Anderson P~W 1958 {\em Phys. Rev.\/} {\bf 109} 1492--1505

\end{thebibliography}

\end{document}